\documentclass[sigconf,authorversion,nonacm]{acmart}

\AtBeginDocument{%
  \providecommand\BibTeX{{%
    \normalfont B\kern-0.5em{\scshape i\kern-0.25em b}\kern-0.8em\TeX}}}

\usepackage[utf8]{inputenc}
\usepackage{multirow}

\usepackage{array}
\usepackage{xcolor}
\usepackage{subfig}
\newcommand{\added}[1]{{#1}}
\newcommand{\final}[1]{{#1}}

\copyrightyear{2022} 
\acmYear{2022} 
\setcopyright{acmlicensed}\acmConference[CHI '22]{CHI Conference on Human Factors in Computing Systems}{April 29-May 5, 2022}{New Orleans, LA, USA}
\acmBooktitle{CHI Conference on Human Factors in Computing Systems (CHI '22), April 29-May 5, 2022, New Orleans, LA, USA}
\acmPrice{15.00}
\acmDOI{10.1145/3491102.3517717}
\acmISBN{978-1-4503-9157-3/22/04}

\begin{document}

\title{A Comparative Evaluation of Interventions Against Misinformation: Augmenting the WHO Checklist}

\author{Hendrik Heuer}
\email{hheuer@uni-bremen.de}
\orcid{0000-0003-1919-9016}
\affiliation{%
 \institution{University of Bremen}
 \city{Bremen}
 \country{Germany}
 }

\author{Elena L. Glassman}
\email{glassman@seas.harvard.edu}
\orcid{0000-0001-5178-3496}
\affiliation{%
 \institution{Harvard University}
 \city{Cambridge, MA}
 \country{USA}
 }

\renewcommand{\shortauthors}{Heuer and Glassman}

\begin{abstract}
During the COVID-19 pandemic, the World Health Organization provided a checklist to help people distinguish between accurate and misinformation. In controlled experiments in the United States and Germany, we investigated the utility of this \final{ordered} checklist and designed an interactive version to lower the cost of acting on checklist items. Across interventions, we observe non-trivial differences in participants' performance in distinguishing accurate and misinformation between the two countries and discuss some possible reasons that may predict the future helpfulness of the checklist in different environments. The checklist item that provides source labels was most frequently followed and was considered most helpful. Based on our empirical findings, we recommend practitioners focus on providing source labels rather than interventions that support readers performing their own fact-checks, \final{even though this recommendation may be influenced by the WHO's chosen order}. We discuss the complexity of providing such source labels and provide design recommendations.
\end{abstract}



\maketitle

\section{Introduction}

Helping users who encounter misinformation online is more important than ever. The COVID-19 pandemic is a prime example of the contemporary abundance of misinformation. At the beginning of the COVID-19 pandemic in 2020, the World Health Organization (WHO), among others, warned that ``misinformation costs lives''~\cite{world_health_organization_managing_2020}. The Office of the Surgeon General of the U.S. Department of Health and Human Services argued that ``limiting the spread of health misinformation is a moral and civic imperative that will require a whole-of-society effort'' \cite{surgeongeneral2021healthmisinfo}. Examples of misinformation in this context included false statements about the origin of the virus, how the virus is spread, how it can be prevented, and how it can be treated as well as false information about its mortality and the vaccines designed to prevent it~\cite{brennen2020types,enders2020different,surgeongeneral2021healthmisinfo}. COVID-19 is, however, only one of the more visible topics for which misinformation is spread. Other topics for which a significant amount of misinformation \final{is} spread include the alleged connection between autism and vaccines \cite{surgeongeneral2021healthmisinfo}, ``AIDS denialism'' \cite{surgeongeneral2021healthmisinfo}, the Pizzagate conspiracy theory~\cite{robb_2020} and QAnon~\cite{phillips_2020}, the latter of which played an important role in the 2021 United States Capitol attack~\cite{doi:10.1177/00113921211034896}.

A large body of research exists on why people are prone to believe in misinformation~\cite{pennycook2020falls,Scheufele7662,altay2020if}, how to study the spread of misinformation~\cite{flintham_falling_2018,pasquetto2020tackling,shu2019studying}, and which approaches may help against misinformation~\cite{lazer_science_2018,potthast2016clickbait,10.1145/3137597.3137600}. There are also a number of tools and interventions against misinformation available to end-users, including reminders that shift users' attention to accuracy~\cite{farnaz2021lightweight,pennycook2021shifting} and novel user interface items like warnings, related articles, and other interface changes~\cite{10.1145/3415211}. %
There is, however, a gap regarding research that compares different approaches regarding their perceived and actual helpfulness, especially for the setting where users review more than the headlines of articles.

There is also data available on individuals' behavior in current information ecosystems, regardless of the availability of particular tools and interventions.
In a 2019 survey of a nationally representative panel of randomly selected U.S. adults conducted by the Pew Research Center, roughly 50\% of respondents believed they sometimes come across ``made-up news and information,'' and nearly four in ten respondents believe they often do~\cite{Mitchell_Pew_2019}. In the same survey, nine out of ten adults believed that made-up news causes a great deal of confusion about the basic facts of current events. Given this information environment, a large majority of respondents reported checking facts in response to what they determine to be possible misinformation, including 77\% of 19-29-year-olds, 81\% of 30-49-year-olds, and 75\% of people aged 50+. The majority of respondents checked facts regardless of their self-rated political awareness.
This motivated us to investigate how to best support people \final{with} fact-checking news and information they encounter online. %

\begin{figure*}
    \centering
    \includegraphics[width=\linewidth]{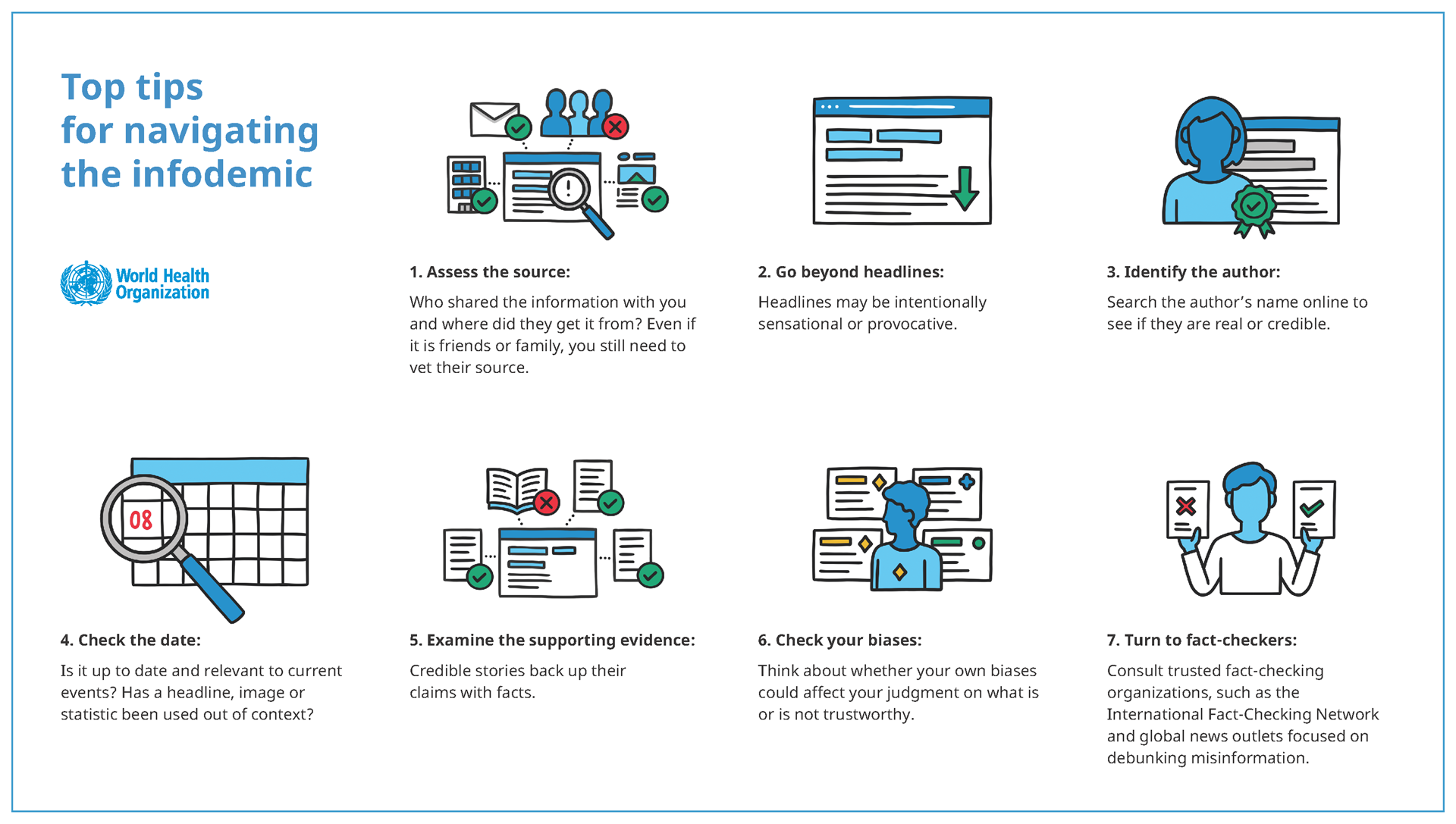}
    \caption{The written checklist that we evaluate in the experiments is based on the World Health Organization's ``Top tips for navigating the infodemic''~\cite{who_2020}.}%
    \label{fig:who}%
    \Description{This figure provides a screenshot of the written checklist that we evaluated in the experiments. This written checklist is based on the World Health Organization's "Top tips for navigating the infodemic.}
\end{figure*}

To design effective interventions against misinformation, it is important to understand what misinformation looks like in practice. A 2020 investigation of COVID-19 misinformation by Brennen et al.~revealed that the majority of COVID-19 misinformation (59\%) includes existing and often true information~\cite{brennen2020types}. This information is then spun, twisted, recontextualized, or reworked. They also find that a substantial amount of misinformation is completely fabricated (38\%). Their analysis of misinformation also showed that 69\% of total social media engagement with misinformation is due to accounts by politicians, celebrities, and other prominent public figures. This is corroborated by a 2021 report by the Center for Countering Digital Hate, which found that twelve individuals are responsible for 73\% of Facebook's anti-vaxx content~\cite{ccdh_disinformation_2021}. 

Tools that empower users are needed because companies like Twitter, Facebook, and YouTube \final{do} not act on known and labeled misinformation~\cite{brennen2020types}. On Twitter, 59\% of debunked COVID-19 misinformation remained available. Every third debunked story (27\%) on YouTube and every fourth debunked story (24\%) remained online as well. This is especially problematic considering the scientific consensus around COVID-19 and the enormous public attention that the pandemic received. This shows that interventions against misinformation that empower individuals are an important and timely topic to study.

This paper responds to the Office of the U.S. Surgeon General's call to ``equip Americans with the tools to identify misinformation''~\cite{surgeongeneral2021healthmisinfo}. In this paper, we investigate two interventions that can be shown to users in situations where they may encounter misinformation. \added{The first intervention is based on the widely publicized checklist shown in Figure~\ref{fig:who}. The checklist was released by the World Health Organization (WHO) to help people navigate misinformation\final{~\cite{who_2020}}}. The checklist includes recommendations to identify the authors, check the date, and examine the supporting evidence, among others. \added{We compare the written checklist to an interactive checklist that we designed and developed. The interactive checklist has the same items as the written checklist, but we augmented these recommendations technically, e.g., by providing source labels and by automatically retrieving information like the headline, the name of the author, and the date on which \final{the} article was published. We also integrated a custom search form that suggests keywords and that automatically searches fact-checking sites. We investigated whether users' performance at recognizing misinformation is improved if following the WHO's recommendations is made less effortful. We examine which recommendations in the checklist users act on and how the different interventions influence users' performance at rating the reliability of news articles. Investigating the written checklist is important because it was published by the World Health Organization, a global authority on public health. Understanding the helpfulness of the recommendations by the WHO is useful because it allows us to provide the public with more effective recommendations. Developing and evaluating the interactive checklist allows us to assess where and how technology can help people recognize misinformation.}

We find that \final{the} interactive checklist is preferred over the written checklist. We also learn that the written checklist \added{is preferred over the control condition without help. \final{For the ordered WHO checklist that we evaluated, we found that the} recommendation to assess the source of a news story is perceived as the most helpful recommendation in both countries. This means that participants perceived it as helpful to know whether a source is seen as reliable or unreliable. The recommendation to assess the source is also the most acted on recommendation. The recommendation to consult fact-checking organizations and to search for information that corroborates or contradicts certain claims was perceived as less helpful, even though we provided a tool that automatically helped users search the fact-checking sites.} The recommendation to fact-check is also acted on less frequently. \final{However, this recommendation was also the component at the end of the checklist, which could have influenced this result.} In Germany, we can see that both the written and interactive checklists are associated with significantly better performance at the task of rating the reliability of news articles though not in the U.S. We discuss possible explanations for this and provide concrete design recommendations on how this can inform interventions that support users to recognize misinformation.

\section{Background}

\added{The goal of our investigation is to help users distinguish reliable from unreliable information. Throughout the paper, we will use the term misinformation to describe all kinds of incorrect or unreliable information. Wardle et al. operationalize misinformation as false information, including false connections and misleading content. They use the term disinformation to describe information that is false and that has an intent to harm~\cite{wardle2018thinking}. Since intent to harm is challenging to prove, we focus on the term misinformation.}

\subsection{People's Proneness to Believe In Misinformation}

Prior research showed that a belief in misinformation is linked to prior exposure~\cite{pennycook2018prior}, how compatible it is to prior beliefs~\cite{doi:10.1177/1529100612451018}, whether a person has a tendency to overclaim one's level of knowledge~\cite{pennycook2020falls}, a person's ability and motivation to spot falsehoods~\cite{Scheufele7662}, the novelty of the false content~\cite{vosoughi2018spread}, and whether a person would find it interesting if the information was true~\cite{altay2020if}. Belief in misinformation is also connected to a tendency to ascribe profundity to randomly generated sentences~\cite{pennycook2020falls}. Other factors that mislead users are a scientific presentation of content, the usefulness of the information, visual design, and an organization's apparent authority~\cite{wineburg2017lateral}. Education, on the other hand, is predictive of a decreased belief in conspiracy theories~\cite{10.1002/acp.3301}. The same is true for analytic thinking, which correlates negatively with believing in fake news~\cite{pennycook2020falls}. 

Research indicates that, on average, people are good at distinguishing the quality of news articles~\cite{pennycook2019crowdsourcing}. At the same time, research also showed that a small number of extreme users struggle with this task, even though the majority is able to reliably rate their trust in news~\cite{10.1145/3240167.3240172}. Researchers also showed that the flagging of misinformation can be outsourced to users~\cite{kim2018leveraging,pennycook2019crowdsourcing,bhuiyan2020investigating,10.1145/3411764.3445433}. 

Pennycook and Rand argued that users' susceptibility to misinformation is better explained by lack of reasoning than by motivated reasoning~\cite{pennycook2019lazy}. This connects to research that shows that people might share misinformation because they are not paying attention or because the social media context \added{distracts} them from assessing the accuracy of information~\cite{pennycook2021shifting}. Other explanations for this include that the role of individual rationality is overstated and that decision-making is strongly influenced by shared group-level narratives~\cite{sloman2018knowledge}. 

Wineburg and McGrew found that professional fact-checkers read \textit{laterally}, i.e., they scan the original online article quickly and then open new browser tabs of additional information to judge the credibility of a website~\cite{wineburg2017lateral}.
In contrast, other participants who were not professional fact-checkers read vertically, i.e., they stayed within the page an online article was on to evaluate its reliability. Overall, Wineburg and McGrew argue that professional ``fact-checkers read less and learned more''. \added{Like} boyd~\cite{boyd_2018_b}, Wineburg and McGrew\final{, therefore,} warn that the wrong kind of media literacy may be taught~\cite{wineburg2017lateral}.

To understand why people believe in misinformation, it is useful to understand why people believe in real news. Jahanbakhsh et al. compiled a taxonomy of reasons why people believe a news claim is true or not~\cite{farnaz2021lightweight}. These factors include having firsthand knowledge, that other trusted sources confirm the claim, that the information is from a source they trust, that the article provides evidence, or that the information is consistent with a user's past experience. Factors why people disbelieve a news story include that they have (firsthand) knowledge on \final{a} topic, that the information contradicts information that a user knows from a trusted source, and that the information is inconsistent with a user's own experience. Users also assess whether a claim is motivated or biased and take into account how the information is presented. In addition to that, users believe that if a piece of information were true, they would have had heard about it. Other aspects that influence belief in news include the logos and the domain name~\cite{wineburg2017lateral} and the number of quoted sources~\cite{sundar1998effect}. 

\subsection{Interventions Against Misinformation}

In this paper, we investigate ways of supporting users in \added{recognizing misinformation. This is a challenging problem because people may be more likely to stick to their initial decisions than to change their opinion, no matter what reasons they are presented with~\cite{stanley2020resistance}. Stanley et al.} describe this phenomenon as a prior-belief bias. This bias connects to related research that showed the influence that motivated reasoning and the alignment of a claim with one’s prior policy position has on the assessment of misinformation~\cite{kahne2017educating}. However, a large body of prior research demonstrated that users consistently and reliably change their beliefs if persuaded~\cite{aldrich1989foreign,rabinowitz1989directional,wood2019elusive}. Another important investigation by Nyhan et al. \added{examined} the impact that journalistic fact-checks of claims made by former U.S. President Donald Trump \added{had} on those who support him and those who do not~\cite{nyhan2020taking}. They find that fact-checks improved the accuracy of users' factual beliefs. At the same time, fact-checks did not affect attitudes towards Trump. Goldberg et al., however, also showed that Republicans in the U.S. shifted their views on climate change after being presented with facts on climate science from trusted messengers~\cite{goldberg2021shifting}.

Several technical solutions have been proposed to automatically detect misinformation using machine learning\final{~(ML)} and data mining techniques~\cite{10.1145/3137597.3137600,10.1145/3305260,potthast2016clickbait,castillo2011information}. Such ML-based approaches try to predict misinformation from lexical-, syntactical-, semantic-, and discourse-level features~\cite{10.1145/3377478}. A large body of these approaches classifies news articles into the categories ``true'' and ``false'' based on the content or meta-data~\cite{perez-rosas-etal-2018-automatic,wang-2017-liar}. However, critical work by Asr et al. found that the available datasets to train ML-based systems are limited and that ensuring data quality is a challenging problem~\cite{doi:10.1177/2053951719843310}. ML-based systems also have the potential to make biased decisions that discriminate against specific groups or individuals~\cite{d2020data, eubanks2018automating, noble2018algorithms}. 

Technical approaches that go beyond predicting whether something is true or false are primarily aimed at researchers or platform providers. Such tools can predict propagation pathways of a message~\cite{10.1145/3159652.3159677,shu2019studying}, detect social bots~\cite{davis2016botornot}, and monitor how misinformation and fact-checks are spread~\cite{10.1145/2872518.2890098}. More socio-technical approaches \added{are} also used, e.g., to support online fact-checkers on the social media website Reddit to identify check-worthy claims using argumentation mining and stance detection~\cite{10.1145/3308560.3316734}, or to nudge users to reflect on the credibility of news they see on Twitter using a browser extension~\cite{bhuiyan2018feedreflect}. %

We follow a socio-technical approach akin to Jahanbakhsh et al., who studied interventions at the moment of sharing misinformation online~\cite{farnaz2021lightweight}. They show that the sharing of false content can be reduced by interventions like asking users to provide an accurate assessment and by asking them to reduce the sharing of false content (even though this also reduces the sharing of reliable information). A socio-technical approach is important because interventions against misinformation can backfire~\cite{greenhill2017rumor,doi:10.1177/1529100612451018,wood2019elusive}. Prior work showed that if users are corrected by experts, the trustworthiness of the news sources shared by a user decreases~\cite{mohsen_2021}. In addition to that, the language toxicity and the partisan slant of users' tweets increase. This indicates important limitations regarding the social correction of misinformation and highlights the importance of performing user studies for any kind of intervention.

\subsection{Warning Labels for Sources \& Content}

The written and interactive checklists that we investigate in this paper relate to research that showed that both false and true headlines are perceived as less accurate when people receive a general warning about misleading information~\cite{ecker2010explicit,clayton2020real,pennycook2020fighting}. A recent study in the context of COVID-19 misinformation showed that reminders to think of the accuracy of a news article can triple the level of truth discernment in users' sharing intentions~\cite{pennycook2020fighting}. However, research also indicates that the specific wording matters and that the improvement is only moderate~\cite{clayton2020real}. This relates to work by Pornpitakpan, who shows that readers are more likely to believe a message from a source with high credibility than a source with low credibility~\cite{pornpitakpan2004persuasiveness}. 

The importance of source assessments is widely recognized~\cite{arnold2021source,10.1145/3415211,swire2017processing,metzger2010social,berinsky2017rumors}. Warning labels were, e.g., shown to reduce users’ intentions to share false news stories on Facebook~\cite{mena2020cleaning}. However, research also showed that even if people see and understand a correction about misinformation, their feelings towards a source may remain unchanged~\cite{swire2020they}. Furthermore, Dias et al.~also found that showing the source of a news article does not affect whether users perceive a headline as accurate or whether they would consider sharing a headline~\cite{dias2020emphasizing}. Dias et al.~also found a strong correlation between trust in a news outlet and the perceived plausibility of a headline. Source labels \final{were also shown to reduce belief in disinformation claims and users' sharing intentions of disinformation~\cite{arnold2021source}}. This, however, depends on partisanship, social media platform, and the specificity of the label. A risk associated with only labeling known misinformation is the so-called ``Implied Truth Effect'' where headlines that are not tagged as false are automatically considered to be validated and are thus seen as more accurate~\cite{pennycook2020implied}. Gao et al.~also warn that labels can have undesirable effects on facilitating the spread of fake news, e.g. by making users look for opinions that they agree with or by making fake news articles \added{appear} more trustworthy~\cite{10.1145/3274324}. Overall, the related work shows the potential of labels. \added{At the same time, an investigation of the perceived helpfulness of such labels for full articles and the effect of labels on people's task performance at rating news articles is missing}.

\added{Investigating the written and the interactive checklist in different countries is important because misinformation is a global problem. The Reuters Digital News Report 2020 indicates that citizens are concerned about misinformation, even though the level of concern varies from country to country~\cite{newman_overview_2020}. For the report, citizens in 40 countries were asked whether they agree with the statement: ``Thinking about online news, I am concerned about what is real and what is fake on the internet''. The concern was the highest in Brazil, where 84\% of respondents are concerned, and the lowest in the Netherlands, where 32\% are concerned. In regards to the two countries that we examined, the United States of America has a comparatively high level of concern (67\%), while Germany has a comparatively low level of concern (37\%). The lack of concern connects to prior research by Humprecht et al., who showed that some countries are more resilient to misinformation than others~\cite{doi:10.1177/1940161219900126}. Motivated by these insights, we investigate one country from the cluster of countries with high resilience to online misinformation (Germany) to a country from the cluster with low resilience (United States).}

\added{Beyond this, the amount of cross-country research is limited, especially for research with a focus on human-computer interaction. Most available research is from political science and psychology. Pennycook et al., for instance, examined the influence of political polarization and motivated reasoning in a cross-country setting including the U.S., Canada, and the United Kingdom~\cite{doi:10.1177/01461672211023652}. Their investigation showed that COVID-19 skepticism in the U.S. is connected to distrust in liberal-leaning mainstream news outlet\final{s; at th}e same time, political conservatism \final{was associated with} misperceptions, e.g., about the risks associated with COVID-19, weaker mitigation behaviors, and a stronger hesitancy to get vaccinated. A related investigation by Stier et al. examined the relationship between populist attitudes and the consumption of various types of online news~\cite{Stier2020}. They \final{measured} media exposure in five countries: France, Germany, Italy, Spain, the United Kingdom, and the United States. They \final{found} that people with populist attitudes \final{consumed} more hyperpartisan news. In Germany, legacy press and public broadcasting \final{were} most frequently visited by participants. In the U.S., commercial broadcasting and digital-born outlets \final{were} most popular. In another striking cross-country investigation, Shirish et al.~examined the impact of mobile connectivity and freedom on fake news propensity during the COVID-19 pandemic~\cite{doi:10.1080/0960085X.2021.1886614}. They \final{found} that nations with more mobile connectivity and more political freedom tend to also have more COVID-19 related misinformation. At the same time, more economic and media freedom \final{was} connected to less COVID-19 misinformation.}

\added{The amount of research that compares interventions in different countries and cultures is limited. This paper addresses this gap by investigating the WHO checklist in two countries: one that has low resilience against misinformation and where citizens are concerned about misinformation (United States) and one country that has high resilience and where citizens are less concerned about it (Germany).}

\added{\section{Checklists against Misinformation}}

\begin{figure}%
  \centering
  \includegraphics[width=\linewidth]{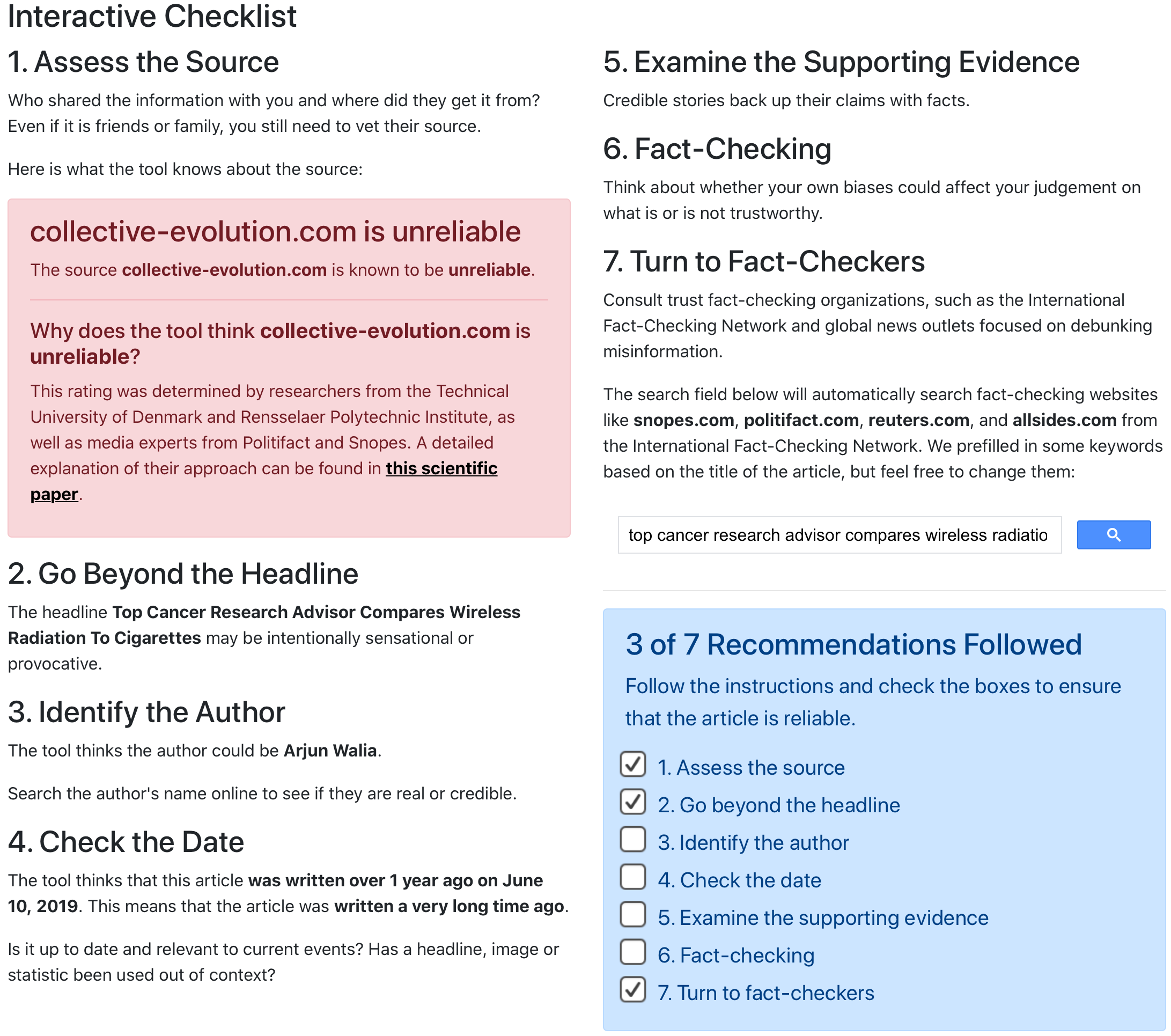}
  \caption{The Interactive Checklist extends the WHO's written checklist by automatically displaying additional information. We augment the checklist items 1.~Assess the Source, 2.~Go Beyond the Headline, 3.~Identify the Author, 4.~Check the Date, and 7.~Turn to Fact-Checkers.}~\label{fig:checklists}
  \Description{This figure provides a screenshot of the Interactive Checklist that we developed. This Interactive Checklist extends the WHO's written checklist by automatically displaying the additional information described in the paper. We augment the checklist items 1. Assess the Source, 2. Go Beyond the Headline, 3. Identify the Author, 4. Check the Date, and 7. Turn to Fact-Checkers.}
\end{figure}

To investigate how to best support users in dealing with misinformation, we took the written checklist provided by the WHO as a starting point\final{~\cite{who_2020}} \added{and conducted the first empirical investigation of its helpfulness}. We then augmented that checklist with a number of \added{technical} tools that we developed to lower the cost of following the checklist items, creating an interactive version of the WHO checklist. \final{The checklist was adapted by the WHO from information compiled by The Spinoff, an online magazine from New Zealand~\cite{who_2020}.}

\added{In the experiment, we} presented participants with three interventions: 1.~the Written Checklist, 2.~the Interactive Checklist, and 3.~the No Help Condition. The Written Checklist (Figure~\ref{fig:who}) is based on the ``Top tips for navigating the infodemic'' \added{that} the World Health Organization published at the beginning of the COVID-19 pandemic in 2020. The checklist includes seven items relevant to identifying misinformation: 1.~Assess the Source, 2.~Go Beyond the Headline, 3.~Identify the Author, 4.~Check the Date, 5.~Examine the Supporting Evidence, 6.~\final{Check Your Biases}, and 7.~Turn to Fact-Checkers. We use the recommendations by the World Health Organization verbatim\footnote{\final{There is a small inconsistency between the checklist published by the WHO and the two checklists that we used in our experiment: the title of the sixth component is mislabeled as ``6.~Fact-Checking'' instead of ``6.~Check Your Biases''; this title for step 6 is still distinguishable from ``7.~Turn to Fact-Checkers'' and the instructions for step 6, which were about checking one's biases, were identical to the original WHO checklist. This mislabeling of step 6 was present for both the Written Checklist and the Interactive Checklist, i.e., this change did not affect the comparison of the interventions. It could, however, have affected the perception of the sixth and seventh component (even though participants were presented with a screenshot of each component in the Component Survey). Subsequently, in the paper, we refer to the bias-checking step by its title, ``6.~Fact-Checking''.}} (but removed the illustrations seen in Figure~\ref{fig:who}).

The Interactive Checklist shown in Figure~\ref{fig:checklists} augments the Written Checklist in a number of ways. We refer to these augmented checklist items as components. The 1.~Assess the Source component displays a source label that indicates whether a source is reliable or unreliable. This source label is also explained. For the U.S.-based experiment, the explanation stated that the rating was determined by researchers from the Technical University of Denmark and Rensselaer Polytechnic Institute, as well as media experts from Politifact and Snopes. Gruppi et al.~describe this meta-ranking in a scientific paper~\cite{gruppi2020nelagt2019}. \added{We provided a link to the paper to the participants.} The explanation of unreliable sources in Germany described that the source had been repeatedly flagged for spreading spam, fake news, or hate speech by a large number of users. For reliable sources, the explanation stated that the website is a well-known newspaper that is widely recognized as reliable. We will explain the labels and explanations in more detail in the Methods section.

For the 2.~Go Beyond the Headline component, we automatically retrieved the headline of the article and displayed it in the context of the recommendation to go beyond the headline. We also automatically retrieved the name of the author and displayed it next to the recommendation (3.~Identify the Author). In addition to that, we contextualized the date by adding additional information like ``This article was written recently'' (within the last three months), ``a while ago'' (within the last year), or ``a very long time ago'' (older than a year) (4.~Check the Date). Since 5.~Examine the Supporting Evidence is based on a close reading of the text and since 6.~Fact-Checking is focused on reflecting on personal biases, we were not able to augment these items. The most elaborate technical support is provided by the 7.~Turn to Fact-Checkers component. For this, we developed a custom Google Search form and automatically pre-filled \final{it with the title of the article}. The search was limited to known fact-checking websites like AllSides, PolitiFact, and Snopes (for the U.S.~experiment) and Correctiv.org, BR Faktenfinder, Tagesschau Faktenfinder, and Mimikama (for the German experiment). The Interactive Checklist also allowed participants to check off which of the recommendations they followed. The Interactive Checklist visualized how many of the components were checked. An animation was shown once all seven recommendations were checked.

In the text that follows, we will use the term ``components'' to refer to both the items of the Written Checklist and the components of the Interactive Checklist. We also compare the perception of the different components of the checklist between the following subgroups: 1.~age (young adults, middle-aged adults, and older adults), 2.~education (those whose education is below or above the sample median), and 3.~political stance (conservatives and liberals).

\section{Methods}

Using the interventions described in the previous section, we conducted two parallel studies to answer the following four research questions:

\begin{itemize}
\item \textbf{RQ1:} Do people perceive either the written or interactive checklists as helpful? How helpful is each checklist component perceived to be?
\item \textbf{RQ2:} Which components of the checklist are acted on? For which components is acting perceived as helpful?
\item \textbf{RQ3:} How does acting on the recommendations impact the performance at rating reliable and unreliable news articles?
\end{itemize}

\added{RQ1 provides the first empirical investigation of the WHO checklist and helps us to understand how technology may help users recognize misinformation.} We hypothesized that users would perceive both the written and the interactive as significantly more helpful than the No Help Condition (RQ1). Informed by prior work on interventions against misinformation, such as~\cite{10.1145/3415211,farnaz2021lightweight}, and work that showed that the role of the source is overrated~\cite{dias2020emphasizing}, we hypothesized that fact-checks could be the most helpful component of the checklist (RQ1) and the component that users acted on the most frequently (RQ2). We also hypothesized that acting on the recommendations would significantly improve users' article ratings across countries, considering related work that suggests that simple reminders can be sufficient to make users more aware (RQ3)~\cite{10.1145/3274324,mena2020cleaning,10.1145/3415211}.

\subsection{Procedure}
\label{sec:proc}
Our goal is to investigate the helpfulness of the different interventions in a realistic setting. \added{Despite the fact that news is increasingly curated by algorithms, e.g., on Facebook~\cite{eslami_i_2015} or YouTube~\cite{10.1145/3415192}, a large proportion of users directly access news websites. The Reuters Institute Digital News Report 2020 indicates that 56\% of males and 44\% of females in Germany and 54\% of males and 46\% of females in the U.S.~directly accessed one or more news website\final{s} or applications~\cite{holig2020reuters}. The proportion of users who do this is similar in size to those who rely on social media to access news: 46\% male, 54\% females in both Germany and the U.S.} We, therefore, asked participants to rate the reliability of different news articles. For each news article, participants rated their agreement with the statement: ``I believe that the information in this news article is reliable.'' on a 5-point Likert scale with the options ``Strongly disagree'', ``Disagree'', ``Neither agree nor disagree'', ``Agree'', and ``Strongly agree''. We use these \final{subjective reliability} ratings to investigate whether the interventions affect the perceived reliability of news articles.

\added{We conducted within-subjects experiments in Germany and the United States. We thus controlled for the variance introduced by participants' backgrounds. A participant's experience with misinformation and different sources was the same across all test conditions. This is crucial because background and experience are hard to control for. They can, however, have a huge influence on people's performance, especially with politically charged issues~\cite{doi:10.1177/1077699018769906}.}

\begin{figure}
    \centering
    \includegraphics[width=\linewidth]{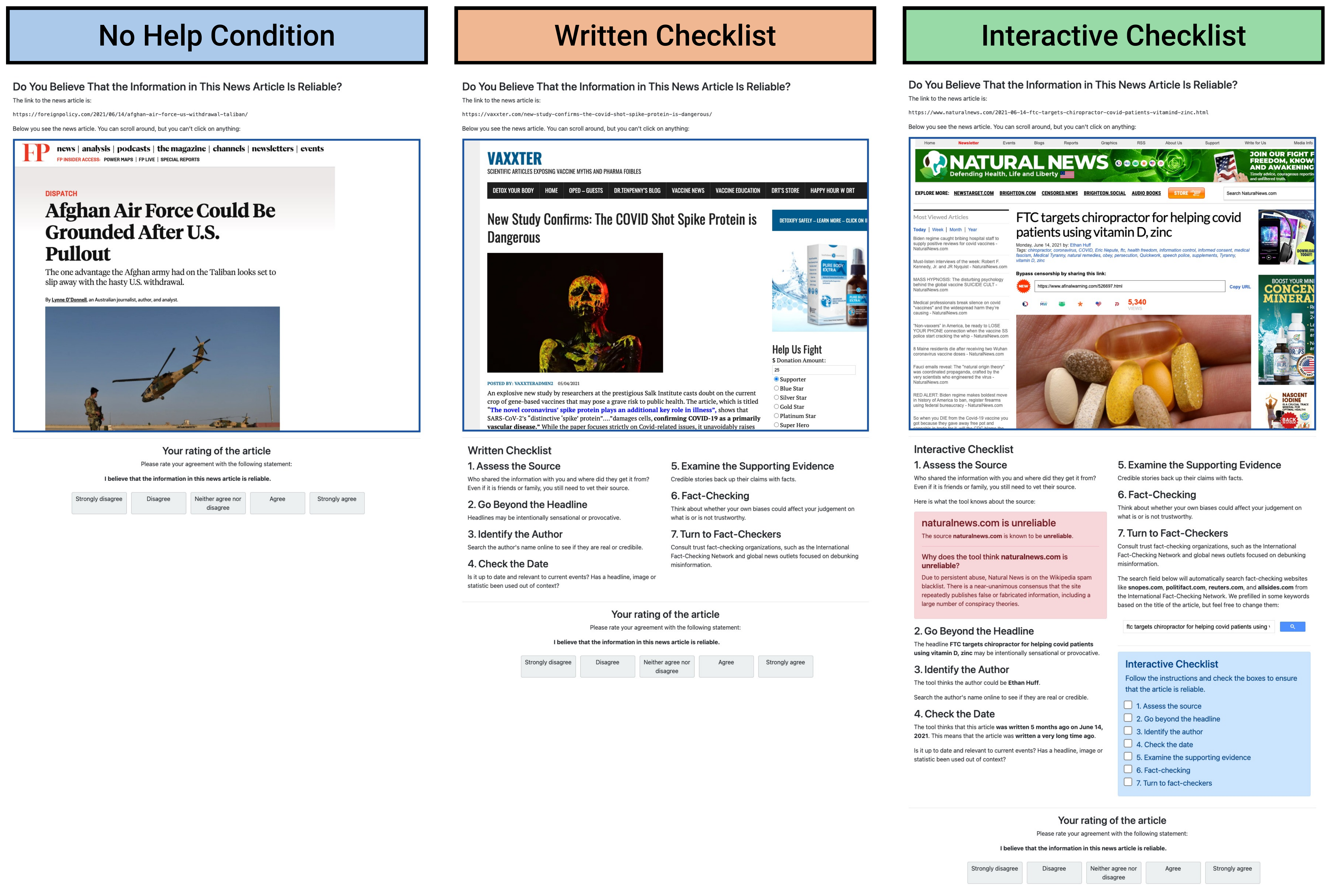}
    \caption{\added{\textbf{Interventions:} Screenshots of the three interfaces that the participants interacted with: the No Help Condition, where users received no support, the Written Checklist, a checklist provided by the World Health Organization (Figure~\ref{fig:who}), and the Interactive Checklist that augments the written checklist. The Interactive Checklist (Figure~\ref{fig:checklists}) provides source labels and automatically retrieves information like the headline, the author, and the date on which an article was published. It also includes a tool that searches fact-checking sites and suggests \final{the title of the article as} keywords. Users were able to freely scroll through the entire article.}}%
    \label{fig:interventions}%
    \Description{This figure depicts the three interventions and how they were presented. We show screenshots of the three interfaces that the participants interacted with: the No Help Condition, where users received no support, the Written Checklist, a checklist provided by the World Health Organization (Figure 1), and the Interactive Checklist that augments the written checklist. The Interactive Checklist (Figure 2) provides source labels and automatically retrieves information like the headline, the author, and the date on which an article was published. It also includes a tool that searches fact-checking sites and suggests keywords. Users were able to freely scroll through the entire article.}
\end{figure}

\begin{figure}
    \centering
    \includegraphics[width=\linewidth]{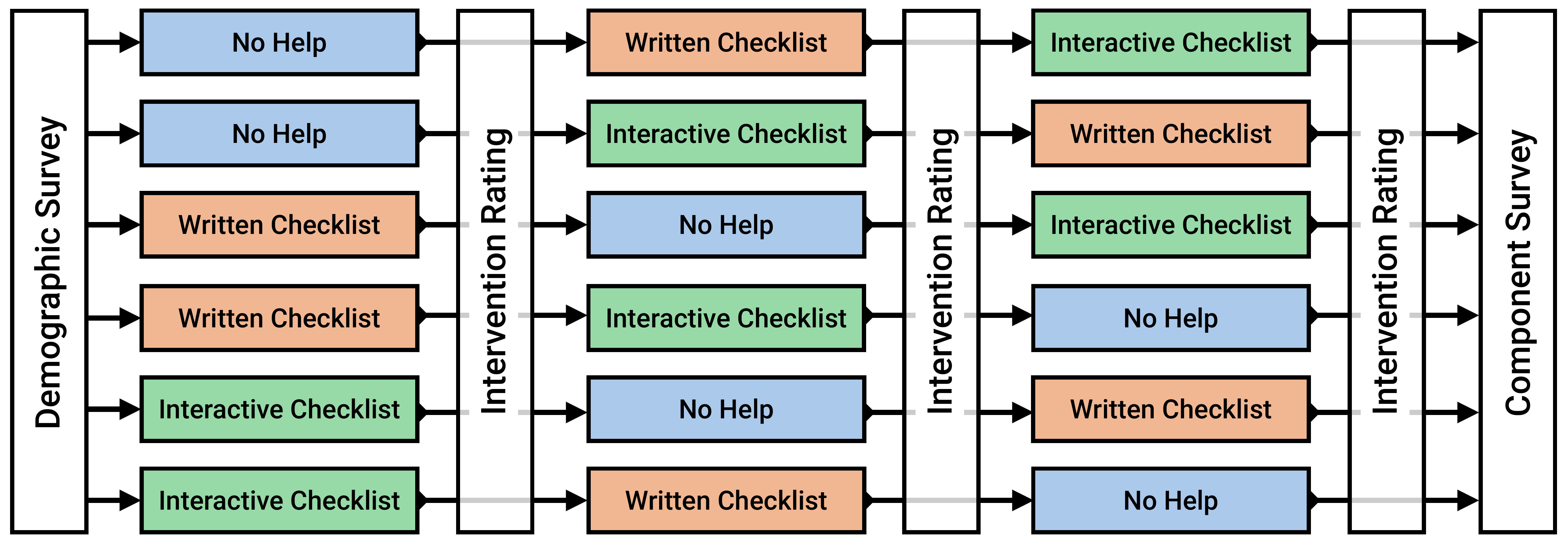}
    \caption{\added{\textbf{Procedure:} Participants rated four articles per intervention. After that, they rated the intervention. The order of the news articles was randomized for each participant. We also randomized the order of the interventions. We excluded the first article per intervention from the analysis to allow participants to familiarize themselves with the intervention. At the end of the study, participants were surveyed about the components of the checklist.}}%
    \label{fig:procedure}%
    \Description{This figure visualizes the procedure of the experiment. Participants rated four articles per intervention. After that, they rated the intervention. The order of the news articles was randomized for each participant. We also randomized the order of the interventions. We excluded the first article per intervention from the analysis to allow participants to familiarize themselves with the intervention. At the end of the study, participants were surveyed about the components of the checklist.}
\end{figure}

\added{For each of the three interventions shown, we presented participants with four randomly selected news articles out of our pool of 12 articles. Figure~\ref{fig:interventions} depicts the three interventions and how they were presented. Users were always able to scroll through the entire article. We visualized the procedure of the experiment in Figure~\ref{fig:procedure}. For each intervention, participants reviewed} three unreliable news articles and one reliable article. We did not want to exclusively present unreliable articles because users might realize that all articles are unreliable and rush through the ratings. At the same time, we wanted to maximize the number of unreliable news stories \added{that} users evaluate. The first news article per intervention was not included in the analysis to allow participants to familiarize themselves with the intervention. Participants were not made aware of this. To limit order effects, we randomized the order of interventions. We also randomized the order of the news articles and we showed them for different interventions. \added{The WHO checklist that we investigated was compiled by an important global authority on public health and it was used during the pandemic to support people, which makes it worth investigating. To understand the efficacy of the WHO checklist, we did not change the order of the checklist components. \final{As described in the Results and Discussion sections, there is a potential order effect regarding the order of the checklist components. We deliberately kept the components in the same order as the WHO to maximize the ecological validity of our findings. This decision could have influenced the results regarding the helpfulness of the individual components. It does not affect the comparison of the Written and the Interactive Checklist.}}

After rating four news articles per intervention, participants evaluated the interventions \added{shown in Figure~\ref{fig:procedure}}. Participants rated their agreement to the statement: ``I felt supported when making my decision''. The question was phrased as follows: ``Please rate your agreement with the following statements in regards to reviewing the last four news stories''. We added ``with the written checklist'' and ``with the interactive checklist'' based on the intervention. The agreement was rated on a 5-point Likert scale. Options included ``Strongly disagree'', ``Disagree'', ``Neither agree nor disagree'', ``Agree'', and ``Strongly agree''. 

After rating all 12 news articles and after evaluating all three interventions, participants were shown a post-experiment survey. Participants were presented with a screenshot of the seven components of the Interactive Checklist. They were then asked to rate their agreement to statements like ``Being reminded of the recommendation \final{`}1.~Assess the Source' was helpful.'' on the 5-point Likert scale that we used before. They also rated their agreement to statements like ``I acted on the recommendation \final{`}1.~Assess the Source' in some way'' with the options ``Yes'' and ``No''.

For the augmented components of the checklist, we also asked participants whether they found it helpful that they acted on the recommendation. The goal was to see if there are differences between the perceived helpfulness of the recommendation from the checklist and the perceived helpfulness of acting on the component. For this, participants had to check the box next to each interface element that they found helpful. The augmented components include: 1.~Assess the Source (which presented information about the reliability of a source), 2.~Go Beyond the Headline (which displayed the automatically extracted headline), 3.~Identify the Author (presenting the automatically extracted author name), 4.~Check the Date (with an automatically extracted and contextualized date), and 7.~Turn to Fact-Checkers (a custom search tool based on Google Search that only indexed fact-checks).

We compared the different ratings of the participants using statistical tests. We relied on non-parametric tests for ordinal data to make no assumptions about the probability distribution of the population~\cite{MacKenzie13}. For the correlated samples of the intervention ratings and the post-experiment surveys, we used the Wilcoxon signed-rank test~\cite{wilcoxon1992individual}. For the independent samples of the article ratings, we relied on the Mann-Whitney U~test~(Wilcoxon Rank test)~\cite{mann1947test,MacKenzie13}. Like the t-tests used for continuous variables, these tests provide p-values that indicate whether statistical differences between ordinal variables exist. \added{We \final{also} compute the correlation between task performance, acting on the components, and the perceived helpfulness of the components using Spearman's rank correlation coefficient. \final{In addition to that, we} ran a linear mixed model analysis using R and the lme4 package to understand whether the interventions affect the ratings provided by participants~\cite{lme4,brown2021introduction}.}

\subsection{Selection of Articles}

Unlike prior work primarily focused on headlines~\cite{clayton2020real,farnaz2021lightweight,stanley2020resistance}, we investigated the interventions using full articles. 
In the following text, we will describe how we sourced the different news stories. For the news stories presented in the U.S., we relied on the labels by Gruppi et al.~\cite{gruppi2020nelagt2019}, who compiled a meta-ranking of source labels from Media Bias/Fact Check (MBFC), Pew Research Center, Wikipedia, OpenSources, AllSides, BuzzFeed News, and Politifact. Misinformation stories were selected from news sources with the label ``conspiracy\_pseudoscience''. Reliable stories, were selected from news sources labeled ``least\_biased''~\cite{gruppi2020nelagt2019}. 

The unreliable articles that we used in the United States reported on alleged concerns about COVID-19 vaccines, misleading representations of how COVID-19 cases are counted, CIA ties to a ``child sex cult'', and the risk of a ``climate lockdown'', among others. We checked these articles and found no basis for any of the claims. Reliable articles covered the careers of Putin's daughters, the future of Afghan Air Forces after the U.S. pullout in Afghanistan, and a comparison of why vaccine passports are more popular in Europe than in the U.S. 

The sources for the German misinformation stories were selected from a dataset of German URLs that are frequently reported by Facebook users~\cite{DVN/TDOAPG_2020}. We selected the 20 most frequently reported domains out of 40,000 verified fake news URLs captured by Facebook (based on a third-party fact check by humans). As this dataset did not provide a complement of least biased news sources, we selected news articles from lesser-known, yet reliable regional newspapers from Germany that sell at least 50,000 copies per issue. Since we sampled participants from all over Germany, we maximized the likelihood that users are not familiar with these regional news sources. In the German study, the unreliable news articles covered the alleged deaths caused by measures against COVID-19, how vitamins can supposedly help against COVID-19, that the proven links between Trump associates and Russian officials are a conspiracy by the CIA, as well as that the Rockefeller family engineered the COVID-19 pandemic to perform a ``great reset'', among other stories. We \added{confirmed} that these articles were not correct. Reliable articles included a story about German vacationers describing a COVID-19 lockdown on the Spanish island Mallorca as being in prison as well as projections of the number of COVID-19 cases.

All articles were retrieved from the homepage of the respective websites on the day before the investigation in June 2020. We presented participants with screenshots of each news article to ensure that all participants saw the exact same article. We used an adblocker to limit the effect of ads and ad personalization. No news source was presented more than once.

\subsection{Participants}

To answer our research questions, we performed two independent studies with different news articles in Germany (188 participants) and the United States of America (208 participants). Participants were sourced from a professional audience platform for market research. We recruited a sample that is diverse in terms of gender, age, political stance, and education. IRB-equivalent approval was sought and granted by the responsible authorities. Informed consent (in line with the European GDPR) was obtained from all participants. In the briefing of the experiment, we told participants that we are computer science researchers who want to build a tool that helps people recognize misinformation, i.e., participants knew upfront that their task was to detect unreliable information. \added{For ethical reasons, we chose to make it transparent from the start that users were about to see unreliable information.} This ensured that the content they reviewed did not negatively influence their lives. This could have led to a priming effect that might have made people more aware of misinformation. We screened out participants from countries other than the U.S. and Germany, people younger than 18, and those who selected "Don't know" for any of the demographic questions. We also excluded people who did not complete the study. \added{We employed a number of attention checks following the recommendations of the market research professionals that we collaborated with. To ensure that people paid attention to the articles, we excluded all participants that took less than five minutes because taking less than five minutes indicates that the users did not read the article. To make sure that people paid attention to all items of the checklist, we also excluded all participants that forgot to check any of the survey questions about the helpfulness of the components in the final survey.}

In the following, we will characterize the participants of the two studies. Both studies have gender-balanced samples. In Germany, 48.4\% identified as female, and 51.6\% identified as male. In the United States, 51.9\% identified as female, and 48.1\% identified as male. In Germany, the median age of participants was 54. The youngest participant was 18, the oldest was 78. The mean age of participants was 49.58 years (SD=15.68). In the United States, the median age was 66.5 years. The youngest participant was 24, the oldest was 80. The mean age was 59.02 \added{years} (SD=16.01). 35\% of participants in Germany stated vocational education as their highest degree, followed by middle school (15\%), high schools (15\%), Master's (13\%), and Bachelor's degrees (11\%). 7\% have a professional qualification, 2\% no formal education, and 2\% a doctoral degree. In the United States, 34\% of participants selected a Bachelor's degree as their highest level of education, followed by high school (21\%), a Master's degree (15\%), vocational education (13\%), and professional education (11\%). 4\% of participants in the U.S. had a doctoral degree, 2\% received no formal education.

We also asked participants about their political preferences. In Germany, the political stance of the different participants closely mirrored an opinion poll about the 2021 German federal election that was conducted one day prior to the experiment, i.e., the participants were representative of the voting intentions of the Germans at the time of the experiment. 
The participants stated their voting preference as follows: 27.13\% CDU/CSU, 17.02\% SPD, 20.21\% GRÜNE, 11.70\% FDP, 6.38\% DIE LINKE, and 9.58\% AfD. This is within 2\% of the projection that we used for our sampling. Our sample in the U.S. is similar to prior work~\cite{farnaz2021lightweight}. 45.67\% of participants identified as Democrat\final{s}, 32.22\% as Republican\final{s}, and 21.63\% as Independent\final{s}. Compared to the results of the 2020 United States presidential election, which took place seven months before our investigation, Independents are overrepresented. 

In addition to directly asking participants which political party best describes their political position, we also asked participants whether they consider themselves to be ``Strongly liberal'', ``Somewhat liberal'', ``Moderate'', ``Somewhat conservative'', or ``Strongly conservative'' on social and economic issues. In Germany, 43\% of participants consider themselves to be liberals in regards to social issues, while 18\% view themselves as conservatives. For economic issues, 33\% consider themselves to be liberals and 16\% consider themselves to be conservatives. In the United States, 41\% consider themselves to be liberals in social issues while 38\% describe themselves as conservatives. For economic questions, 34\% \added{regard} themselves as liberals and 40\% as conservatives.  We took the average of these two self-assessments as an indicator of whether a participant is a conservative or a liberal. In Germany, 83 participants are liberals (44\%) and 40 are conservatives (21\%). In the United States, 82 participants are liberals (39\%) and 85 are conservatives (41\%).

\section{Results}

In this paper, we investigate whether people perceive a written and an interactive checklists and their individual components as helpful (RQ1).
We \added{then} examine which components users self-reported acting on (RQ2) \added{and} analyze how acting on the interventions affects article ratings (RQ3).

\subsection{Helpfulness of Checklists (RQ1)}

\begin{figure*}%
  \centering
  \begin{minipage}[b]{0.3\textwidth}
    \centering
    \includegraphics[width=\linewidth]{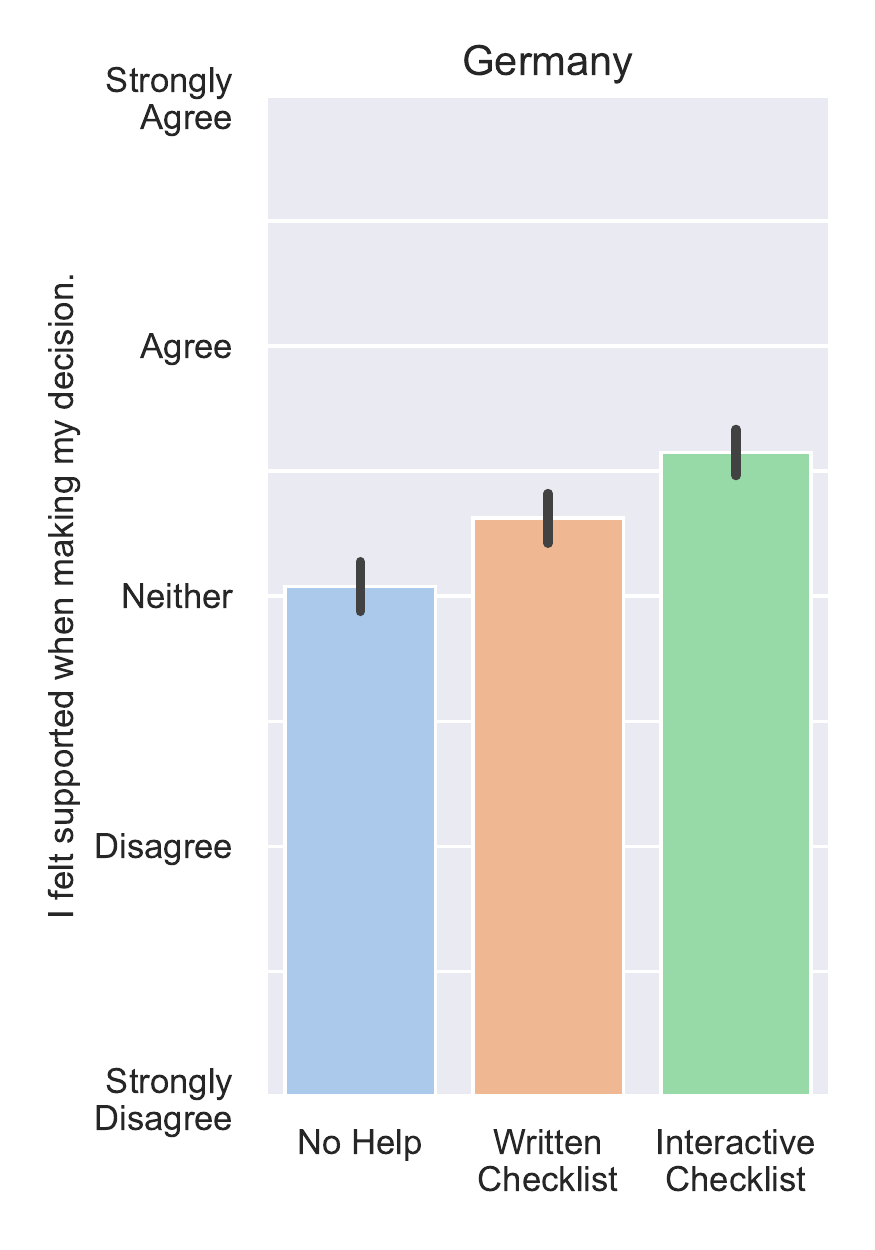}
  \end{minipage}
  \begin{minipage}[b]{0.3\textwidth}
    \centering
    \includegraphics[width=\linewidth]{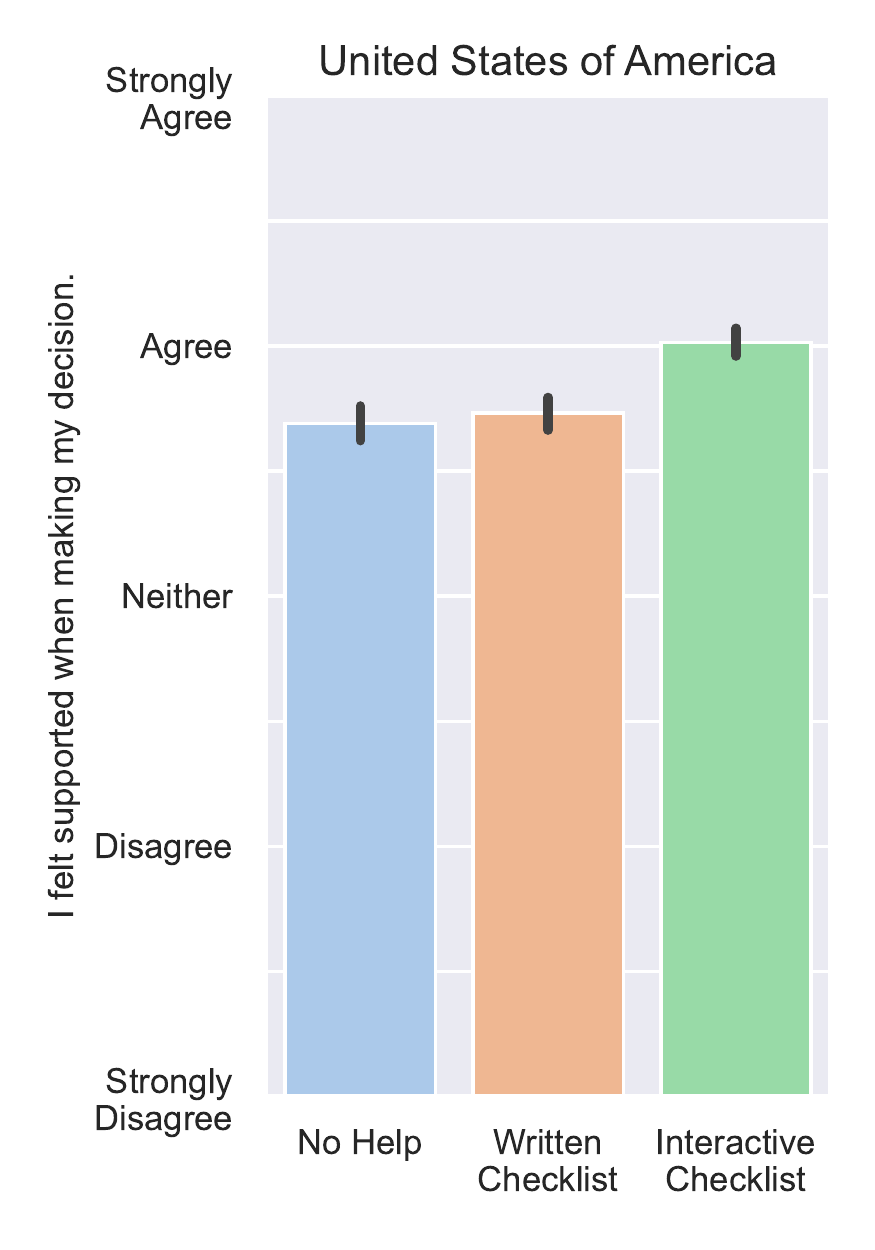}
  \end{minipage}
  \caption{\added{Ratings of the perceived helpfulness of the No Help, Written Checklist, and Interactive Checklist conditions show that the perceived support is the highest for the Interactive Checklist.}}~\label{fig:intervention_ratings}
  \Description{This figure provides a comparison of the perceived helpfulness of the No Help Condition, the Written Checklist, and the Interactive Checklist. The figure shows that the Interactive Checklist is perceived as the most helpful and the No Help Condition is perceived as least helpful in both countries. This can be observed both in the proportion of people who Agree or Strongly Agree that the interventions are helpful. This increases from the No Help Condition to the Interactive Checklist. At the same time, the proportion of people who Disagree or Strongly Disagree that the interventions are helpful decreases.}
\end{figure*}

Figure~\ref{fig:intervention_ratings} shows the perceived helpfulness ratings of the three different interventions. Across both countries, we can observe a monotonic increase in the perceived helpfulness, i.e., a decrease in ``Strongly Disagree'' and ``Disagree'' ratings and an increase in ``Strongly Agree'' and ``Agree'' ratings, as we look from the No Help Condition to the Interactive Checklist. 
We also visually inspected all the subgroups and observe that the Interactive Checklist is perceived as more helpful than the Written Checklist, which is perceived as more helpful than the No Help Condition. 
Aggregated across subgroups in Germany, the No Help condition was perceived as helpful by 35\% of participants. This increased to 44\% for the Written Checklist and to 59\% for the Interactive Checklist. \added{On the 5-point Likert scale, the mean improves from the No Help condition to the Written Checklist. The highest rating can be observed for the Interactive Checklist.} In the U.S., the No Help Condition is perceived as helpful by 60\% of participants. This increased to 64\% for the Written Checklist and 78\% for the Interactive Checklist. \added{Participants in the U.S. gave similar ratings for the No Help Condition and the Written Checklist. The Interactive Checklist has the highest ratings}. A Wilcoxon Signed-Ranks Test indicated that the perceived helpfulness of the Written Checklist and the Interactive Checklist differ significantly from the No Help Condition in Germany (Z=10047.0, p<.001 and Z=8229.0, p<.001, respectively). In the U.S., the Interactive Checklist is also rated significantly different from the No Help Condition (Z=3951.0, p<.001), but not the Written Checklist (Z=10537.5, p=.082). We also found significant differences between the ratings of the perceived helpfulness of the Written and Interactive Checklists in Germany (Z=9240.0,~p<.001) and the U.S. (Z=5482.5,~p<.001).

Within subgroups, our analysis found that the Interactive Checklist \final{is} most popular among young adults (DE: 69\%, US: 100\%). Middle-aged participants perceived it as slightly less helpful (DE: 57\%, US: 92\%). In Germany, we also find that the Written Checklist \final{is} perceived more favorably by educated people (53\%) than by less educated people (37\%). In the United States, this trend \final{is} reversed: less educated people rated the \final{Written Checklist} higher (78\%) than more educated people (56\%). Overall, liberals tended to agree more strongly that the checklists were helpful, both in Germany and the United States. Especially in the U.S., both conservatives (74\%) and liberals (78\%) rate the Interactive Checklist highly. In Germany, liberals (62\%) were more likely to agree that the Interactive Checklist \final{is} helpful than conservatives (50\%). In the United States, this difference is more even pronounced for the Written Checklist (Liberals: 71\%, Conservatives: 58\%). In Germany, the difference is smaller (Liberals: 45\%, Conservatives: 37\%).

\begin{figure*}%
  \centering
  \begin{minipage}[b]{0.49\textwidth}
    \centering
    \includegraphics[width=\linewidth]{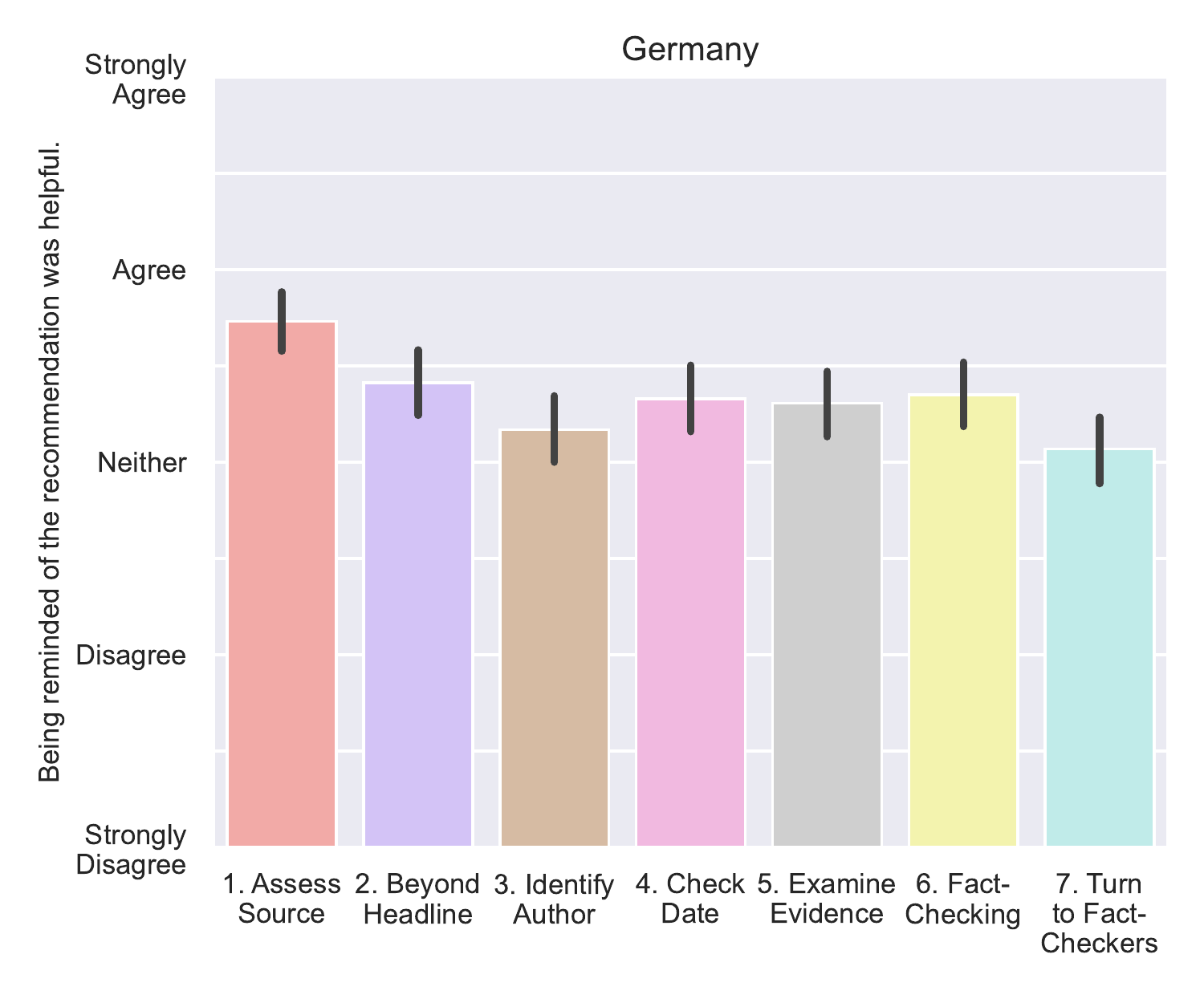}
  \end{minipage}
  \begin{minipage}[b]{0.49\textwidth}
    \centering
    \includegraphics[width=\linewidth]{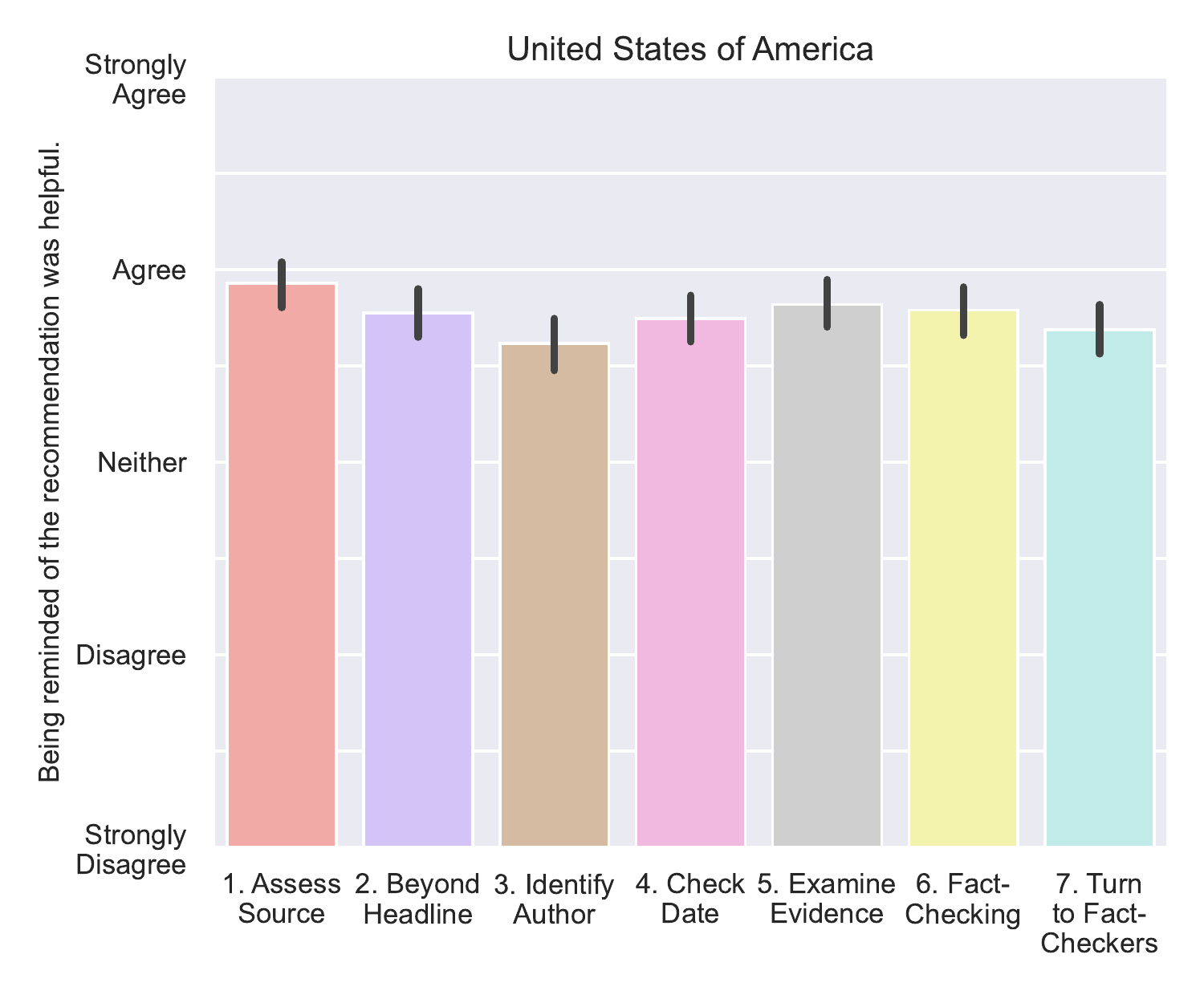}
  \end{minipage}
  \caption{\added{After using the three interventions to rate 12 news articles in total, participants rated which of the components of the checklists they perceived as helpful.}}~\label{fig:recommendations_helpful}
  \Description{This figure visualizes disagreement and agreement to the statement "Being reminded of the recommendation was helpful" for the seven components of the checklists. We find that 1. Assess the Source is perceived as the most helpful component in both countries.}
\end{figure*}

We also investigated the perceived helpfulness of the different components \final{of the Interactive Checklist} (Figure~\ref{fig:recommendations_helpful}). We find that Assessing the Source is perceived as the most helpful component. 64\% of Germans and 76\% of U.S. respondents think that being reminded of this component \final{is} helpful. The second most \final{perceived-to-be} helpful component in both countries \final{is} Going Beyond the Headline (DE: 51\%, US: 69\%). In Germany, only these two components \final{are} perceived as helpful by a majority. The third most \final{perceived-to-be} helpful component in Germany is Checking the Date (47\%), followed by Examining the Supporting Evidence (46\%), Fact-Checking (45\%), and Identifying the Author (40\%). The least \final{perceived-to-be} helpful component \final{is} Turning to Fact-Checkers (37\%). In the United States, the third most helpful component after Assessing the Source (76\%) and Going Beyond the Headline (69\%) is Examining the Supporting Evidence (68\%). This component is followed by Fact-Checking (66\%) and Checking the Date (64\%). Turning to Fact-Checkers (61\%) and Identifying the Author (58\%) \final{are} perceived as the least helpful components.%

The Wilcoxon Signed-Ranks tests show that the component Access the Source is rated significantly higher than all other components in both countries (DE: p<.001, US: p<.05). In Germany, the ratings for the component Turn to Fact-Checkers is rated significantly lower than all other components (p<.05), except for Identifying the Author. In the United States, the ratings for Access the Source and Fact-Checking are also significantly different from all other ratings. The Identify the Author component is also rated significantly different than most other components, except for Examine the Supporting Evidence and Turn to Fact-Checkers in both countries. 

\final{Our results represent how the components were perceived in the context of the WHO checklist. It is important to remember that the order of the components and the way they were presented could have influenced the perception of the different components and their perceived helpfulness.}

\subsection{Acting on Checklist (RQ2)}

\begin{figure*}%
  \centering
  \begin{minipage}[b]{0.49\textwidth}
    \centering
    \includegraphics[width=\linewidth]{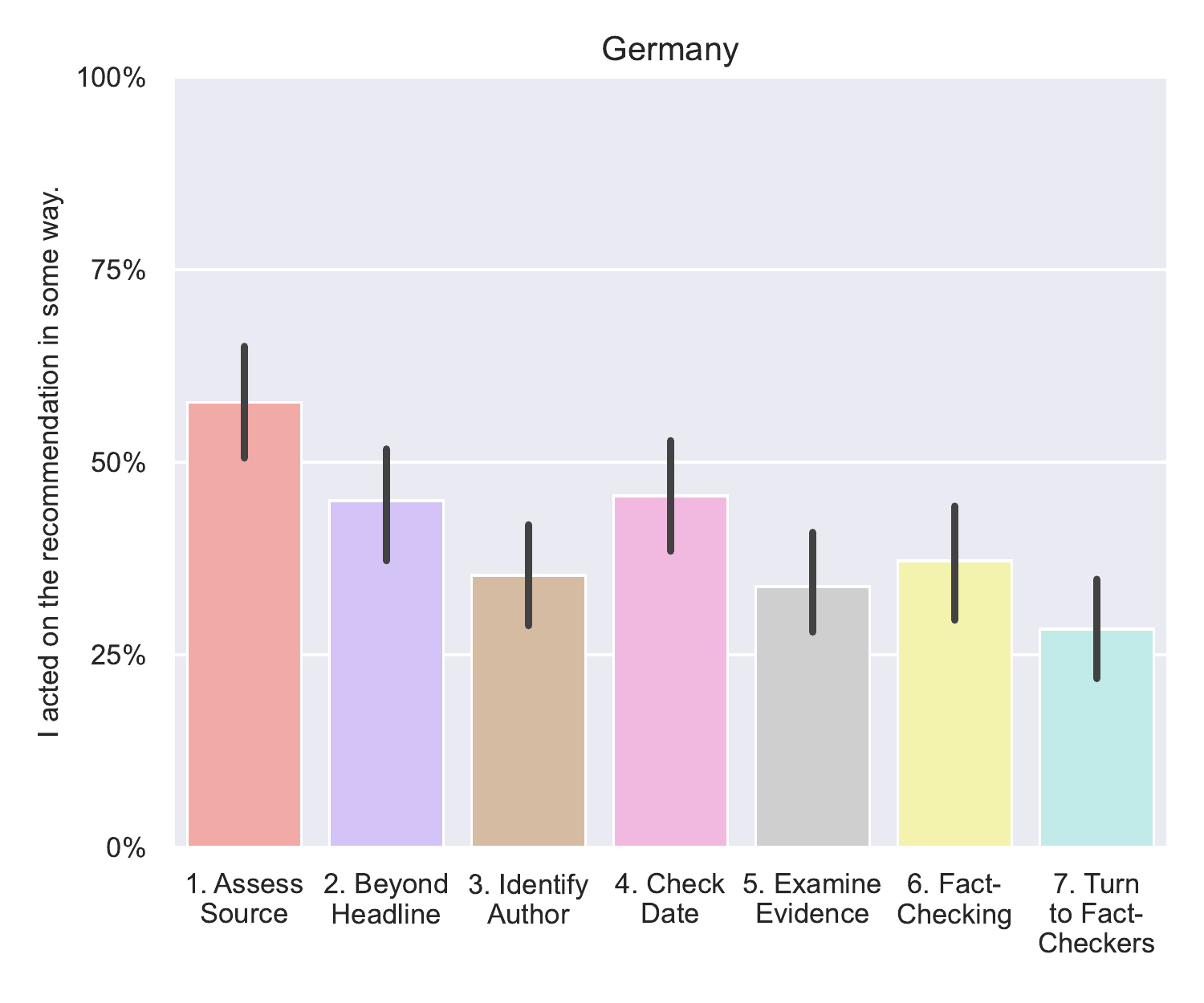}
  \end{minipage}
  \begin{minipage}[b]{0.49\textwidth}
    \centering
    \includegraphics[width=\linewidth]{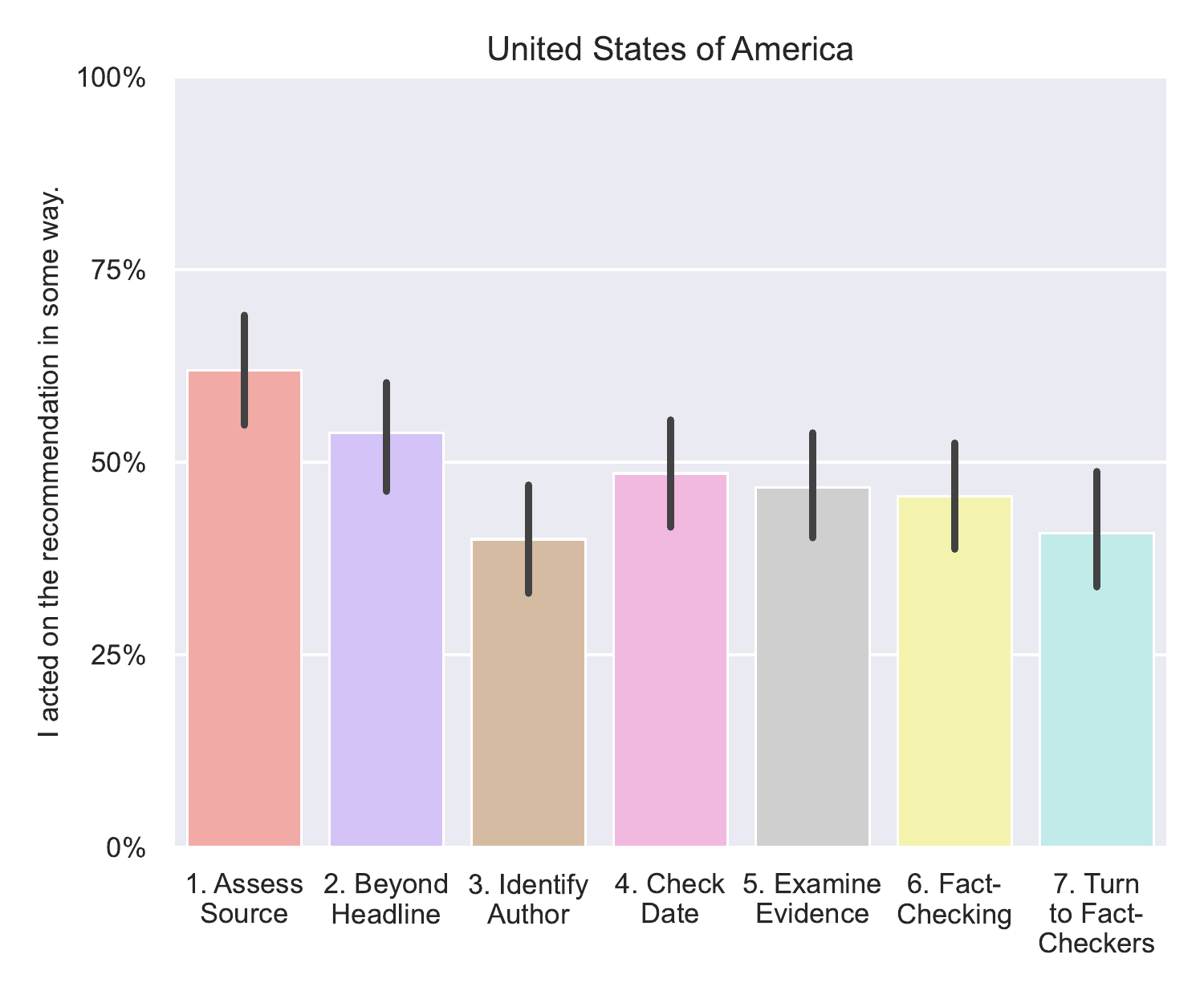}
  \end{minipage}
  \caption{\added{In addition to the perceived helpfulness shown in Figure~\ref{fig:recommendations_helpful}, participants also self-reported which of the components \final{of the Interactive Checklist} they acted on. Only 1. Assess the Source was acted on by a majority in both countries.}}~\label{fig:recommendations_acted}
  \Description{This figure visualizes agreement to the statement "I acted on the recommendation in some way" for the seven components of the checklists. We find that 1. Assess the Source is the most acted on in both countries.}
\end{figure*}

\begin{figure*}%
  \centering
  \begin{minipage}[b]{0.49\textwidth}
    \centering
    \includegraphics[width=\linewidth]{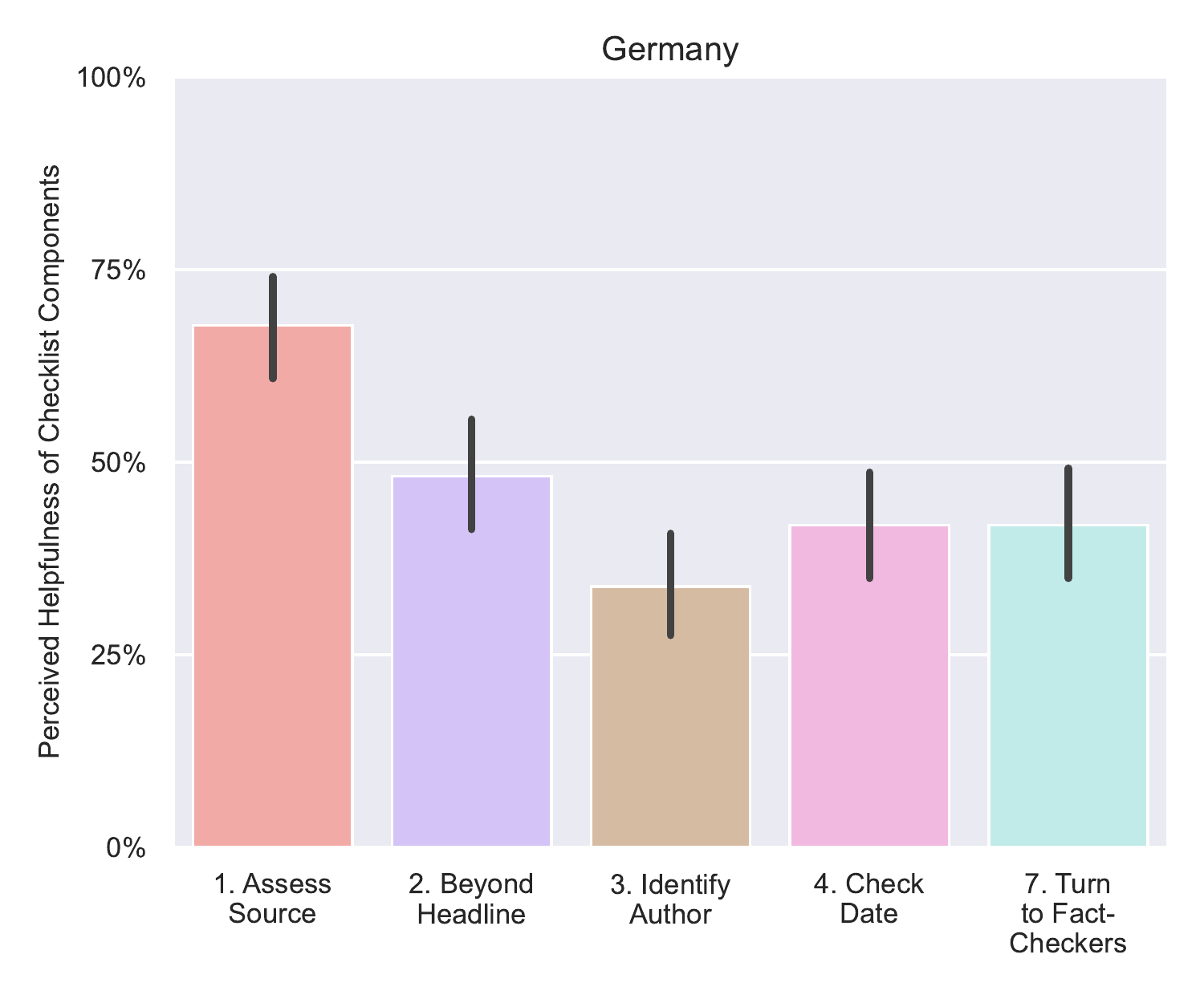}
  \end{minipage}
  \begin{minipage}[b]{0.49\textwidth}
    \centering
    \includegraphics[width=\linewidth]{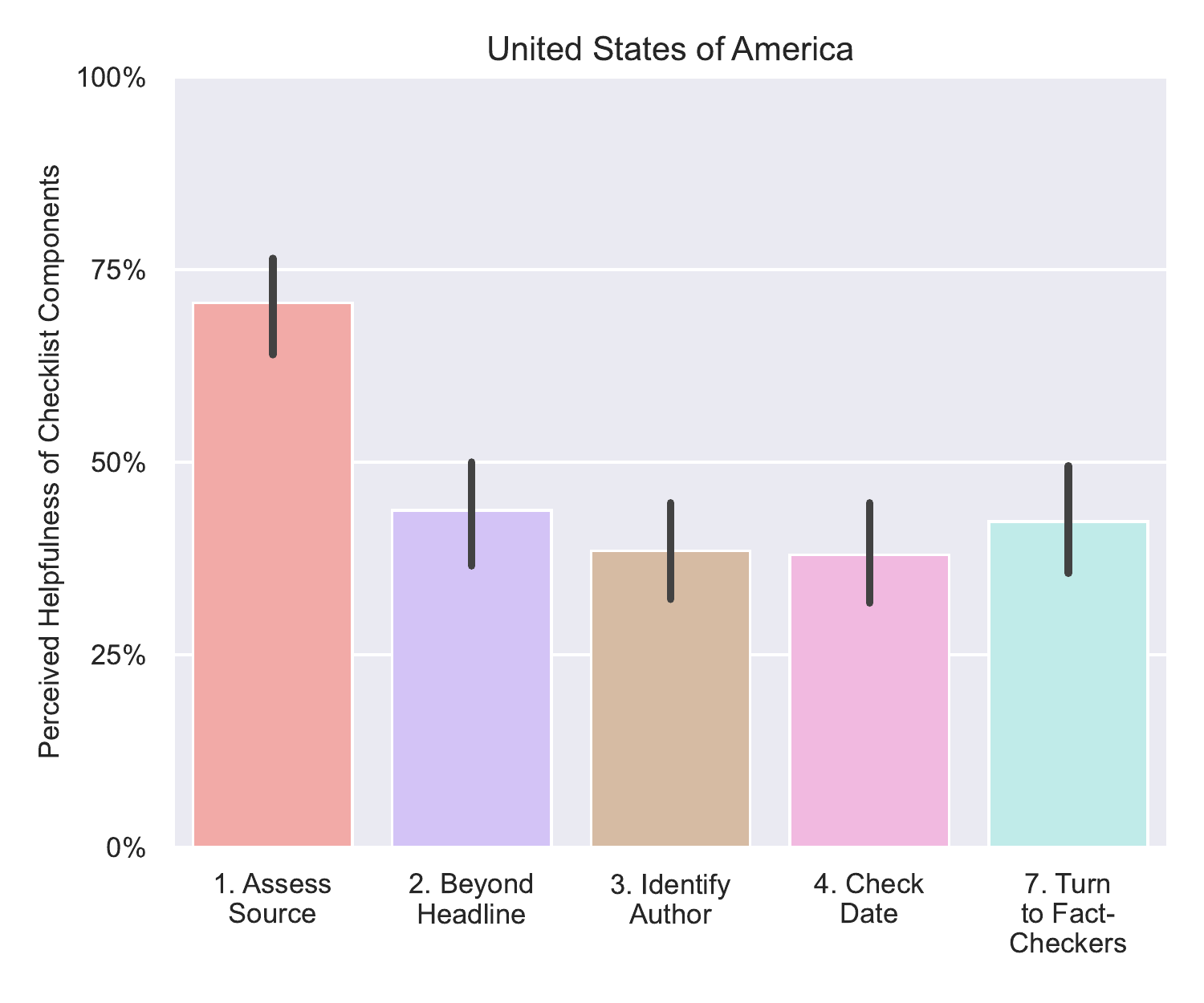}
  \end{minipage}
  \caption{\added{Participants also reported how helpful they found acting on the \final{five augmented} components of the WHO checklist. Only 1. Assess the Source is perceived as helpful by a majority of users.}}~\label{fig:recommendations_usefulness}
  \Description{This figure shows the perceived helpfulness of the five augmented checklist components. Here, again, the 1. Assess the Source component is perceived as the most helpful in both countries.}
\end{figure*}

In addition to surveying people about the perceived helpfulness of the different interventions (RQ1), we also examined which of the components users reported acting on. Figure~\ref{fig:recommendations_acted} \final{shows} users' agreement to the statement ``I acted on the recommendation in some way.'' The Assessing the Source component is the one that most people acted on (DE: 58\%, US: 62\%). In Germany, Checking the Date \final{is} the second most frequently acted on component (46\%), followed by Going Beyond the Headline (45\%). In the United States, Going Beyond the Headline \final{is} the second most frequently acted upon component (54\%), followed by Checking the Date (49\%). In Germany, Turning to Fact-Checkers is the least frequently acted on the component (28\%). In the U.S., it is second to last (41\%) before Identifying the Author (40\%). According to the Wilcoxon Signed-Ranks Test, the frequency with which users engaged with the components differed significantly for all countries and all components (p < .05), with the exception of the ratings of Identify the Author and Examine the Supporting Evidence.

Participants also rated how useful acting on the different components \final{is} (Figure~\ref{fig:recommendations_usefulness}). As explained in Section~\ref{sec:proc}, we only asked this question for components that were augmented (excluding the recommendations to examine the supporting evidence and reflecting on their own biases). In both experiments, acting on the Assessing the Source component \final{is} perceived as most useful (DE: 68\%, US: 71\%). Like with the perceived helpfulness, the Go Beyond the Headline component \final{is} perceived as the second most useful component (DE: 48\%, US: 54\%). The Turn to Fact-Checkers component \final{is} considered to be useful by 42\% Germans and 42\% Americans. This is surprising considering the limited helpfulness of the reminder and the limited number of people that acted on the recommendation. The component for which acting \final{is} perceived as least helpful \final{is} Identifying the Authors (DE: 34\%, US: 38\%). In the U.S., this component is tied with Checking the Date (38\%). Regarding the helpfulness of the components, we again find that the component Access the Source is rated significantly different from all other components according to Wilcoxon Signed-Ranks tests in both countries (DE: p<.001, US: p<.05). In Germany, the Identify the Author component is also distinguished from the Go Beyond the Headline and the Check the Date components.

\final{Here, again, it is important to keep the potential influence of the order in which the WHO arranged the components in mind.}

\subsection{Checklists \& Task Performance (RQ3)}
\label{sec:checklist_task_performance}
\begin{table*}
\small
  \caption{\added{A linear mixed model analysis shows that the Interactive Checklist has a strong positive effect on whether participants provided correct article ratings. We also find a moderate effect of education and age.}}
  \Description{This table presents a linear mixed model analysis that indicates that the Interactive Checklist has a strong positive effect on whether participants provided correct article ratings. We found significant effects for the Interactive Checklist (Estimate ± SE: 0.260 ± 0.093, z = 2.797, p = .005). We also found that education (Estimate ± SE: 0.089 ± 0.034, z = 2.594, p = .009) and age (Estimate ± SE: 0.018 ± 0.004, z = 4.753, p = .000) have a significant influence on participants' performance at the task. *~$\leq$~.05, **~$\leq$~.01, ***~$\leq$~.001.}
  \label{tab:linear_model}
  \begin{tabular}{lrrrl}
\toprule
Predictor & Estimate & SE & z-value & Pr(>|z|) \\
\midrule
Intercept                     &   -1.160 &      .351 &  -3.303 &              .001 \final{***} \\
Intervention (Written Checklist vs. No Help)         &    .046&      .093&   .499&              .618 \\
Intervention (Interactive Checklist vs. No Help) &    .260 &      .093 &   2.797 &              .005 \final{**} \\
Education                       &    .089 &      .034 &   2.594 &              .009 \final{**} \\
Political Stance                &   -.055&      .042&  -1.328&              .184 \\
Age                             &    .018 &      .004 &   4.753 &              .000 \final{***} \\
Country (U.S.)                      &   -.631&      .287&  -2.203&              .028 \final{*} \\
\bottomrule
\end{tabular}
\end{table*}

In addition to the self-reported helpfulness (RQ1) and how useful they found acting on the components (RQ2), we also examined the effect of the Written Checklist and the Interactive Checklist on users' accuracy at rating news articles. 

\final{First, we wanted to see if participants in either country could distinguish between reliable and unreliable articles, in any condition. For each country and for each condition within that country, we ran a Mann-Whitney U test. In both countries, this indicated significant differences between the 5-point subjective reliability ratings that participants gave to reliable and unreliable news articles in the No Help Condition (DE: U=19402.5, p<.001, US: U= 25165.0, p<.001), the Written Checklist (DE: U=15909.0, p<.001, US: U= 27971.5, p<.001), and the Interactive Checklist (DE: U=16290.5, p<.001, US: U= 25731.5, p<.001). The Mann-Whitney U tests showed that the subjective reliability ratings of articles are significantly different.}

\added{To \final{get a more nuanced understanding of what impacts whether a participant can provide} correct article ratings, we ran a linear mixed model analysis using R (4.1.2) and the lme4 package~\cite{lme4,brown2021introduction}. \final{The linear mixed model analysis enabled us to explore what factors influence correct ratings. This analysis allowed us to account for variation in people's baseline levels of reporting and the differences between the news articles.} \final{We included the information on whether a participant provided the correct answer as the dependent variable. For a reliable article, if a participant rated their agreement with the subjective reliability as ``Agree'' or ``Strongly agree'', we counted this as a correct rating. For an unreliable article, if the participant rated their agreement with the reliability statement via the ``Disagree'' or ``Strongly disagree'' options, we counted this as a correct rating. If the participant selected the ``Neither agree nor disagree'' option, we always counted this as an incorrect rating. We} added random effects of the news article that was rated and the participant that rated the tool. We added fixed effects of a participant's education, political stance, age, and country. The model was specified as follows: $correct\_answer \thicksim 1 + intervention + education + political\_stance\\+ age + country + (1 | participant) + (1 | article)$.}

\added{The coefficients of the model are shown in Table~\ref{tab:linear_model}. We found significant effects for the Interactive Checklist vs. No Help (Estimate~±~SE: 0.260~±~0.093, z~=~2.797, p~=~.005), but not for the Written Checklist vs. No Help. We also found that education (Estimate~±~SE: 0.089~±~0.034, z~=~2.594, p~=~.009) and age (Estimate~±~SE: 0.018~±~0.004, z~=~4.753, p~=~.000) have a significant influence on participants' performance on the task, while political stance did not have a significant influence on on participants' performance on the task.}

\added{A likelihood-ratio test indicated that the model including the intervention rating provided a better fit for the data than a model without it, $\chi^2(2) = 8.7895, p=0.012$. An additional comparison between the Interactive Checklist and the Written Checklist yielded a significant advantage for the Interactive Checklist (Estimate~±~SE: 0.213~±~0.093, z~=~2.298, p~=~.021). This means that the Interactive Checklist is significantly better than the Written Checklist.}

\added{We further investigated whether it made a difference if a participant used the intervention for a news article that was from a reliable source or from an unreliable source. For this, we examined the interaction effect of $intervention$ and $article\_rating$. Unfortunately, this model did not converge. We also examined models with random slopes, which did not converge either. We, therefore, excluded these models from our analysis.}

\begin{table}[]
\small
\centering
\caption{\added{The table shows Spearman's $\rho$ for the correlation between correct answers and whether those who provided the answers acted on the different components (Act, Figure~\ref{fig:recommendations_acted}) and whether the participants perceived the components as helpful (Help, Figure~\ref{fig:recommendations_usefulness}). The correlations corroborate the finding that assessing the source is helpful in Germany. *~$\leq$~.05, **~$\leq$~.01, ***~$\leq$~.001.}}
\label{tab:correlations_task_performance}
\Description{This table shows Spearman's Rho for the correlation between correct answers and whether those who provided the answers acted on the different components and whether the participants perceived the components as helpful. We report two important comparisons for the two countries: the correlation between correct ratings and acting on a component (top) and the correlation between correct ratings and perceiving a component as helpful (bottom). The top of the table shows the correlation between correct article ratings and whether the participant that provided the article rating acted on a particular component. In Germany, we find a positive correlation between correct article ratings and acting on the 1. Assess the Source component, r(1600) = .11, p = .000. In the German study, we also see a positive correlation between task performance and acting on the 3. Identify the Author component, r(1636) = .07, p = .003. In the United States, we find no significant correlations between acting on the components and task performance.}
\begin{tabular}{l|ll}
\toprule
$corr(correct\_answer, x)$ & DE & US \\
\midrule
Act: Assess Source & .11  *** & .02 \\
Act: Beyond Headline & .03 & -.01 \\
Act: Identify Author & .07  ** & -.03 \\
Act: Check Date & .03 & .01 \\
Act: Evidence & .05  * & .02 \\
Act: Fact-Checking & .05  * & -.03 \\
Act: Turn to Fact-Checkers & .03 & -.03 \\
\midrule
Help: Assess Source & .13  *** & .07  ** \\
Help: Beyond Headline & .01 & -.02 \\
Help: Identify Author & .03 & -.05 \\
Help: Check Date & .05  * & .02 \\
Help: Turn to Fact-Checkers & .03 & .01 \\
\bottomrule
\end{tabular}
\end{table}

\subsection{Correlation Analysis}
\label{sec:correlation_analysis}
\final{We also analyzed the correlations (1) between task performance and acting on a component, (2) between task performance and perceived helpfulness, (3) between acting on different components, and (4) between perceiving different components as helpful.}

\subsubsection{\final{Correlations Between Task Performance and Acting on a Component}}

\added{To understand the effect of the different components on task performance, we computed Spearman's~$\rho$. In Table~\ref{tab:correlations_task_performance}, we report two important comparisons for the two countries: the correlation between correct ratings and acting on a component (top) and the correlation between correct ratings and perceiving a component as helpful (bottom). The top of Table~\ref{tab:correlations_task_performance} shows the correlation between correct article ratings and whether the participant that provided the article rating acted on a particular component. In Germany, we find a positive correlation between correct article ratings and acting on the 1.~Assess the Source component, r(1600)~=~.11, p~=~.000. In the German study, we also see a positive correlation between task performance and acting on the 3.~Identify the Author component, r(1636)~=~.07, p~=~.003. In the United States, we find no significant correlations between acting on the components and task performance. The differences in the dimensions of freedom are due to the fact that we only compare article ratings of participants that explicitly rated that they did or did not act on the components. We excluded those that skipped the question.}

\subsubsection{\final{Correlations Between Task Performance and \final{Perceived Helpfulness}}}

\added{The bottom of Table~\ref{tab:correlations_task_performance} reports Spearman's $\rho$ for the correlation between correct article ratings and participants' ratings of the helpfulness of the checklist components. In this section, we compare the final surveys of participants to each other. We find that the 1.~Assess the Source component is positively correlated with task performance in both Germany, r(1681)~=~.13, p~=~.000, and the United States, r(1870)~=~.07,~p=.004.}

\subsubsection{\final{Correlations Between Acting on Different Components}}

\added{We also investigated how acting on one component correlates with acting on another component. \final{In the Appendix, we provide Table~\ref{tab:correlations_germany}, which shows all these correlations in Germany, and Table~\ref{tab:correlations_usa}, which shows all the correlations in the United States.} In both, Germany and the United States, we find that acting on one component is correlated with acting on other components. We observe the strongest correlation between those who acted on the 5.~Examine the Supporting Evidence and the 6.~Fact-Checking components (DE: r(180)~=~.57, p~=~.000; US: r(196)~=~.55, p~=~.000). The correlation between acting on the 1.~Assessing the Source and 7.~Turning to Fact-Checkers component is the smallest in Germany, r(178)~=~.19, p~=~.011. In the United States, this correlation is also smaller than other correlations, r(193)~=~.29, p~=~.000.}

\subsubsection{\final{Correlations Between Perceiving Different Components As Helpful}}

\added{To better understand the helpfulness of the different components, we also investigated the correlation between acting on a component and the perceived helpfulness of the component. \final{We report all correlations in the bottom right of Table~\ref{tab:correlations_germany}~(Germany) and Table~\ref{tab:correlations_usa}~(United States) in the Appendix}. In Germany, we find a correlation between acting on the 1.~Assessing the Source component and perceiving the component as helpful, r(178)~=~.34, p~=~.000. The same is true for the 2.~Going Beyond the Headline, r(178)~=~.21, p~=~.005, 3.~Identify the Author, r(182)~=~.33, p~=~.000, 4.~Checking the Date, r(180)~=~.33, p~=~.000, and 7.~Turn to Fact-Checkers components, r(185)~=.18, p~=~.012. Correlations between acting on a component and perceiving a component as helpful can also be observed in the United States. Here, the 3.~Identify the Author component stands out, r(198)~=~.37, p~=~.000. We also identified such correlations for the other components: 1.~Assess the Source, r(195)~=~.23, p~=~.001, 2.~Going Beyond the Headline, r(197)~=~.17, p~=~.018, 4.~Checking the Date, r(200)~=~.17, p~=~.018, and the 7.~Turn to Fact-Checkers, r(199)~=~22, p~=~.002.}

\added{Another striking finding of our analysis is that acting on any of the components is correlated with perceiving the 1.~Assess the Source component as helpful. In Germany, this is also true for the 3.~Identify the Author component. This could imply that acting on the different components could have helped participants to realize that the 1.~Assess the Source component is the most helpful.}

\subsubsection{\final{Correlations Between Acting on a Component and Perceived Helpfulness}}

\added{In the bottom left of Table~\ref{tab:correlations_germany} (Germany) and Table~\ref{tab:correlations_usa} (United States) \final{in the Appendix}, we report the correlations between acting on a component and the perceived helpfulness of the component. In Germany, the strongest correlation for this was between 2.~Going Beyond the Headline and 7.~Turning to Fact-Checkers, r(187)~=~.41, p~=~.000. In the United States, this correlation was much smaller, r(206)~=~.19, p~=~.007. Here, the strongest correlation was between perceiving 4.~Checking the Date and 7.~Turning to Fact-Checkers as helpful, r(206)~=~.33, p~=~.000.}

\subsection{\added{Number of Correct Ratings per Participant}}

\begin{figure}%
  \centering
  \begin{minipage}[b]{0.23\textwidth}
    \centering
    \includegraphics[width=\linewidth]{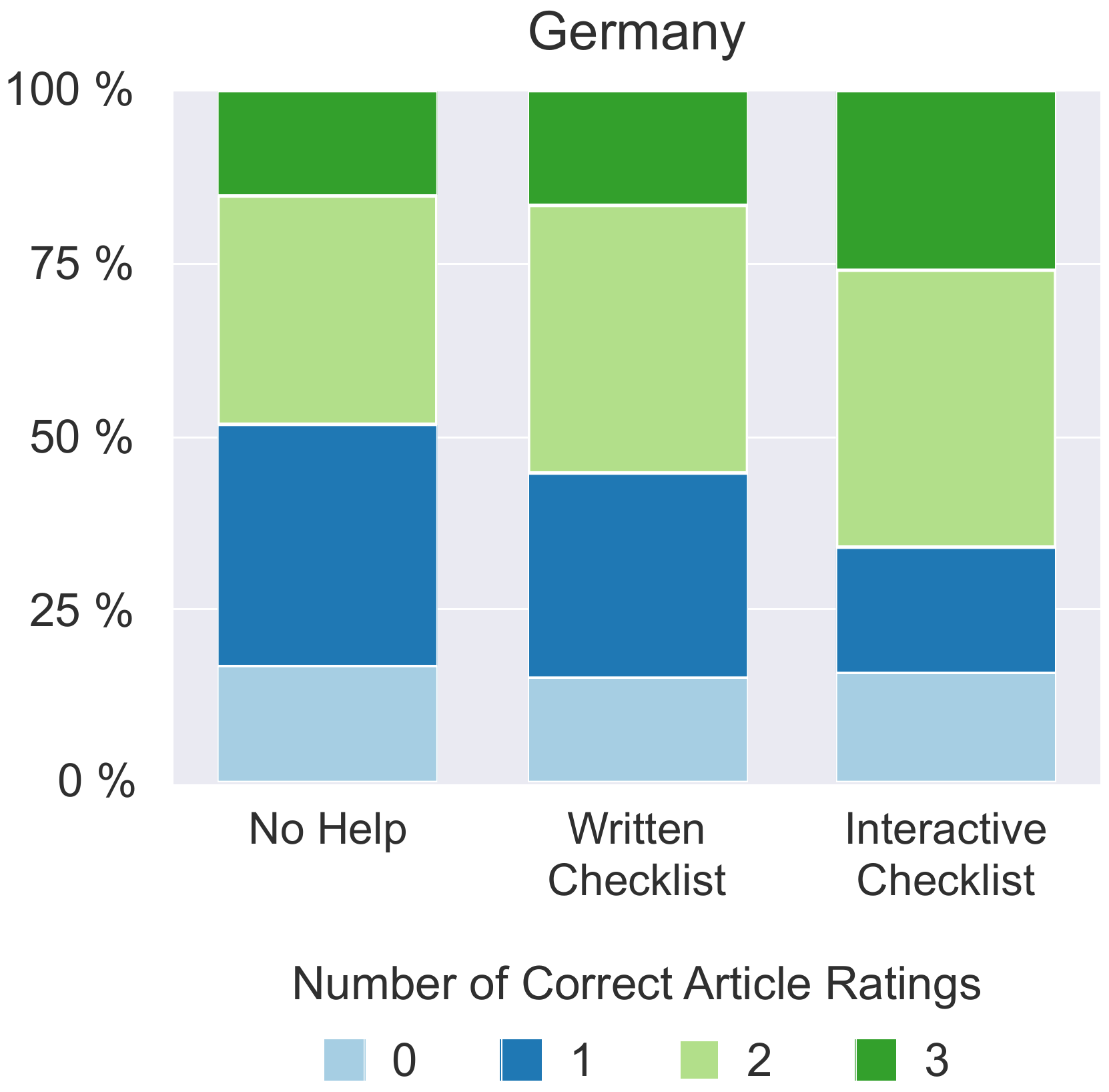}
  \end{minipage}
  \begin{minipage}[b]{0.23\textwidth}
    \centering
    \includegraphics[width=\linewidth]{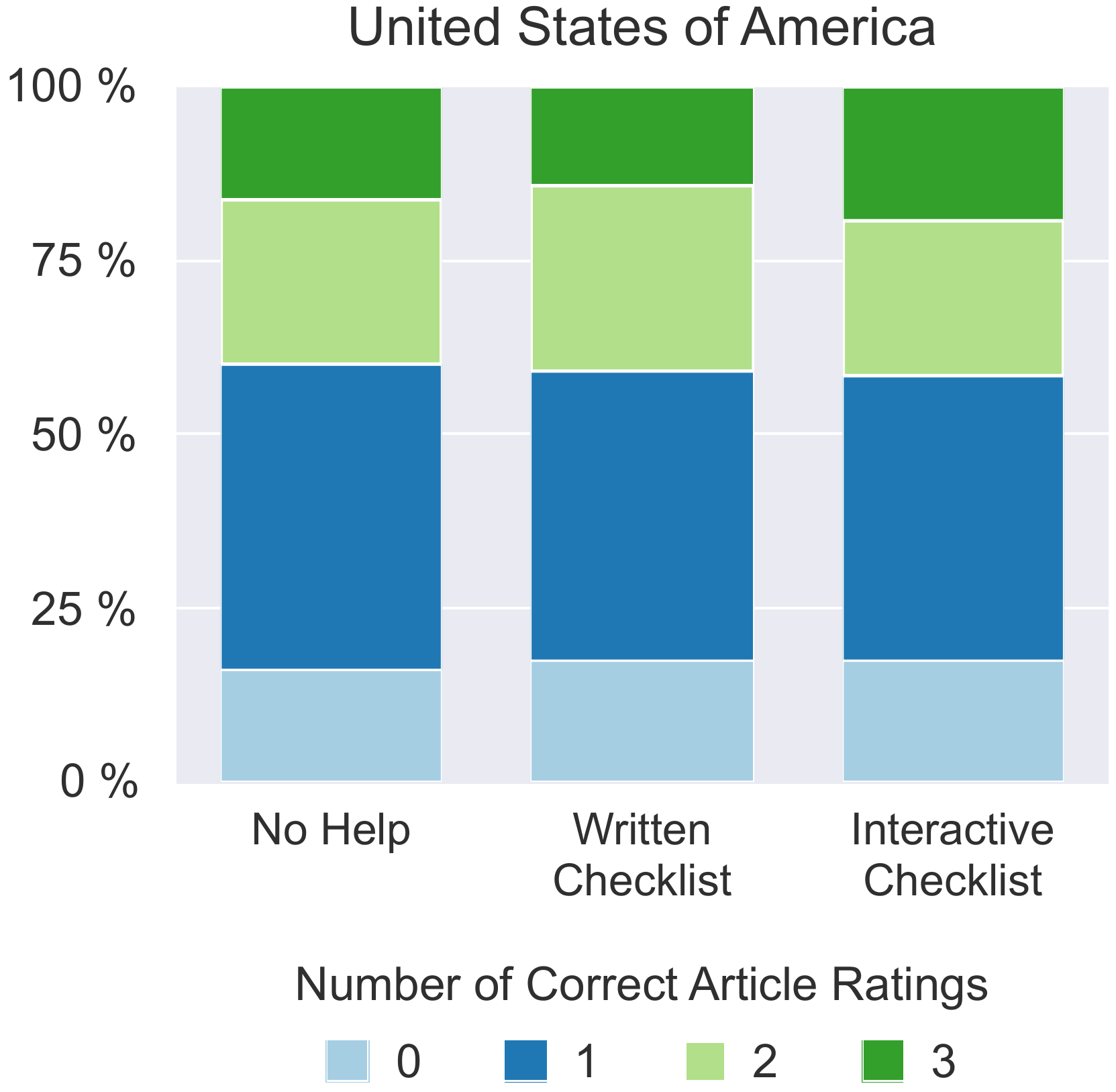}
  \end{minipage}
  \caption{\added{This figure compares how many of the reliable and unreliable news articles are correctly rated for the three conditions that we compared. In Germany, we find that the Written Checklist and the Interactive Checklist led to better article ratings. In the United States, the interventions have no such effect on the ratings.}}~\label{fig:correct_article_ratings}
  \Description{In this figure, we compare how many of the reliable and unreliable news articles are correctly rated for the three conditions that we compared. The results show that especially in Germany, the Written Checklist and the Interactive Checklist lead to better article ratings, i.e. the proportion of participants who rate two or three of the three article ratings correctly increases. In the United States, the interventions have no such effect on the ratings, i.e. the proportion of correct ratings does not increase.}
\end{figure}

\final{The previous two sections showed that (1) the Interactive Checklist has a strong positive effect on providing correct article ratings, i.e., fitting a linear mixed model to examine what influences correct article ratings (Section~\ref{sec:checklist_task_performance}), and (2) that individual checklist components are correlated with providing correct ratings (Section~\ref{sec:correlation_analysis}). As a final step of our analysis, we wanted to examine how the different interventions impact the \textit{number} of correct article ratings each participant provides.}

\added{\final{We visualized how the number of correctly rated articles per participant changes between interventions (Figure~\ref{fig:correct_article_ratings})}}. In Germany, the proportion of participants who rated all three articles correctly increased significantly from 15\% for the No Help Condition and 16\% for the Written Checklist to 26\% for the Interactive Checklist. While 48\% of participants were able to rate two or three out of three articles correctly in the No Help Condition, 55\% are able to do so with the Written Checklist and 66\% were able to do so with the Interactive Checklist. Meanwhile, the proportion of German participants without any correct ratings remained relatively stable across interventions, from 17\% for the No Help Condition to 15\% for the Written Checklist and 16\% for the Interactive Checklist.

Surprisingly, this strong positive effect of the Interactive Checklist cannot be observed in the U.S. The number of participants able to correctly rate all three articles changed only moderately (No Help Condition: 16\%, Written Checklist: 14\%, Interactive Checklist: 19\%). The same applies to the group of those who rate two or more articles correctly (No Help Condition: 40\%, Written Checklist: 41\%, Interactive Checklist: 42\%). The proportion of those who did not get any rating right remains stable as well (No Help Condition: 16\%, Written Checklist: 18\%, Interactive Checklist: 18\%).

We also examined the rating differences between subgroups. A surprising result is how poorly young adults performed in the U.S. With the Interactive Checklist, none of the young adults get two or three ratings correctly. With the No Help Condition and the Written Checklist, only 8\% get two ratings right. Nobody in the young adults' group got three ratings right. This is a significant difference from other age groups in the U.S. Even in the No Help Condition, 20\% of middle-aged or 51\% of elderly people get two or more ratings right. In Germany, young adults (42\%) also performed worse than middle-aged (55\%) or elderly (48\%) people based on their performance at two or three ratings, at least in the No Help Condition, but the difference is not as noteworthy as it is in the U.S.

In regards to education, we can observe some differences in Germany. Without help, 56\% of highly educated people, but only 41\% of less-educated people got two or three ratings correct. With the Interactive Checklist, 62\% of highly educated people and 59\% less educated people got two or three ratings right. In the U.S., such large differences between the No Help Condition and the Interactive Checklist cannot be observed. Without help, 34\% of highly educated and 39\% of less-educated people rated two or three ratings correctly. With the Interactive Checklist, 39\% of educated people and 41\% of less-educated people rated two or three articles correctly.

\final{The linear mixed model analysis presented in Section~\ref{sec:checklist_task_performance} showed that education, age, and country have a significant effect on whether participants provide correct article ratings; political stance \textit{did not}.}
In both countries, we do not observe a difference between conservatives and liberals in the No Help Condition: In Germany, 50\% of liberals and 50\% of conservatives get two or three ratings right. In the U.S., 40\% of liberals and 40\% of conservatives achieve the same result. Liberals improved their performance with the Written Checklist, though these differences were not statistically significant: In Germany, 63\% of liberals rated two or three article ratings correctly, but only 50\% of conservatives did. In the U.S., 46\% of liberals have two or three \final{correct ratings}, but only 34\% of conservatives did. The same is true for the Interactive Checklist, though, again, these differences were not statistically significant: In Germany, 71\% of liberals but only 53\% of conservatives provide two or three correct ratings. In the U.S., 50\% of liberals get two or three ratings right. Only 34\% of conservatives have two or three correct ratings with the Interactive Checklist. 
\final{In summary, while we observed no difference between liberals and conservatives without tools, we started to see a (non-statistically significant) difference when the Written and Interactive Checklists were provided. We report these non-statistically significant differences in case these differences are found to be significant in future work and because our goal is to design tools and interventions that help everyone.}

\section{Discussion}

In this paper, we examined \added{the efficacy of the WHO checklist and the differences between a written checklist and an interactive checklist. We also investigated} which components of the checklist are perceived as helpful (RQ1), which of these components participants act on (RQ2), and how acting on these components affects the performance at rating news articles (RQ3). \added{The most important outcome of this investigation is that users can be supported in the fight against misinformation. We find that the recommendations that the World Health Organization released at the beginning of the COVID-19 pandemic are indeed perceived as helpful by participants, especially when they are supported through technology~(RQ1)}. Our results show that an interactive version of the checklist is perceived as more helpful than the written checklist \added{The linear mixed model analysis provides evidence that the Interactive Checklist also has an effect on whether participants are able to provide the correct answer. Other factors that influence task performance include the education of participants and their age.}

\added{Our findings are particularly notable because the Interactive Checklist encourages the ``lateral reading'' that was shown to significantly improve users' ability to determine the reliability of information~\cite{wineburg2017lateral,breakstone2021lateral}. When evaluating the reliability of information online, Wineburg and McGrew distinguish between those who read vertically and those who read laterally \cite{wineburg2017lateral}. Vertical reading means staying within a website to evaluate its reliability. Lateral reading means quickly scanning a website and opening up new browser tabs to judge the credibility of the original site. Wineburg and McGrew show that those who read laterally make better decisions and take less time. The Interactive Checklist provides shortcuts for actions that Wineburg and McGrew associate with ``lateral reading''. For instance, the Assess the Source component provides an assessment of whether a source is reliable or not. The component collects evidence from a number of websites and online forums. Since the Interactive Checklist provides this information in the user interface, the effort for users is reduced. The same is true for the Turn to Fact-Checkers component. Here again, the laborious process of visiting different fact-checking websites is replaced by a custom search engine that searches relevant fact-checking websites. Using the components of the Interactive Checklist may, therefore,} qualify as ``lateral reading'' because the outcome is very similar to the outcome that actions commonly associated with ``lateral reading'' would produce.
Rather than teaching users how to search and navigate the Internet efficiently, the Interactive Checklist, e.g., directly provides an assessment of the source and search results based on a custom search engine of reliable fact-checking sources. However, despite the documented benefit of ``lateral reading'', we found that especially the custom search engine that provided fact-checks via a custom Google Search form was not perceived as helpful by the participants in our two experiments.

In the following, we will discuss the significance of these results, especially in regards to the benefit of source labels and the limits of fact-checking. We will also address the potential users of the interventions and the apparent complexity of distinguishing reliable from unreliable information. We discuss reasons why the checklist has a noticeable effect in Germany, but not in the United States and provide concrete design recommendations.

\subsection{\final{Order Effects}}

\final{We controlled for order effects between the interventions by randomizing the order in which the different interventions were shown. We also randomized the order in which the participants reviewed the articles. In the Written and Interactive Checklists, we presented the different components in the order published by the WHO. The participants might, therefore, have mistaken the order in which the components were presented in the checklist by the World Health Organization as a ranking of the importance of the components. This could have influenced the ratings of the different components.}

\subsection{Turning to Fact-Checkers}

\final{For the ordered checklist that we investigated, we found surprising differences between how fact-checks and source labels are perceived by participants.}
The \final{technically} most sophisticated 7.~Turn to Fact-Checkers component was perceived as less helpful and was acted on far less than simpler components like 1.~Assess the Source. \added{We also found that turning to fact-checkers did not improve participants' ratings. Our analysis of the correlation between acting on the 7.~Turn to Fact-Checkers component and providing correct ratings did not find an effect.} 

\final{In the context of the ordered checklist of the WHO, we find that the} perceived and actual helpfulness of the fact-checks was limited, and there are a variety of potential explanations for this. Users may have perceived the recommendation as less helpful because they might already know that fact-checkers can help. This, however, does not explain why the recommendation is not acted on. The disparity could be explained by the social desirability bias~\cite{edwards1958social}. Users may simply feel that it is appropriate to agree with the statement that fact-checks are helpful (even though they would not actually fact-check). \final{Another explanation for this could be that the component was perceived as less helpful because it was presented as the last item in the checklist. The result could also be explained by user fatigue. Users might not have engaged with the seventh component in as much detail as with the first.} Another \final{possible reason} is that fact-checking requires a lot of time and effort. A user has to deeply engage with an article to be able to find and comprehend fact-checking information. This could entail a significant time commitment and require a lot of cognitive effort from the user. The practices related to fact-checking are also very different from the ways in which news is consumed. Liu et al., e.g., report that 80\% of the 200,000+ articles that they investigated are visited for less than 70 seconds~\cite{liu2010understanding}. This implies that only a minority reads the whole article. This usage pattern is very different from the deep engagement that fact-checking requires. 

\added{Finally, another possible explanation for this limited helpfulness is the fact that} for fact-checks to be helpful, articles that present an opposing point of view need to exist. Our dataset did, for instance, include the misinformation article ``Secret Bat Lab: Does the US pay for [t]he death of Citizens of Georgia?''. At the time of our investigation, the fact-checking websites that we included did not provide articles that opposed this accusation. The same was true for many other articles included in our study. Even though we leveraged Google's state-of-the-art search engine, the custom search engine did not retrieve articles that presented opposing points of view. One reason for this could be that there is no incentive to write and publish an article that explains why the U.S. does not pay for the death of citizens of Georgia. This could, however, imply that the recommendation to consult trusted fact-checking organizations may not be helpful in a large number of cases. We invite other researchers to systematically investigate the assumption that fact-checks that provide opposing points of view exist.

\subsection{Source Labels}

\final{For the WHO checklist that we investigated, we} found that source labels and tools that help users assess the source of a news article were more helpful than fact-checks. \added{The 1.~Assess the Source component is perceived as helpful, it is acted on the most, and acting on the component is perceived as helpful. Our correlation analysis also indicates that acting on the 1.~Assess the Source component had a strong effect on task performance in Germany.} \final{This could have been influenced by the way that the components were ordered by the WHO. There are, however, other properties that make assessing a source noteworthy.} First and foremost, parsing source labels requires significantly less time than skimming or reading an article or performing a fact-check. Second, in addition to the time requirement, the cognitive effort associated with understanding a source label could also be significantly lower than that for alternatives like turning to fact-checkers. Third, the source labels can be easily integrated into existing user interfaces, both by those who operate a platform as well as by providers of third-party browser extensions. Fourth, there are existing and established jury processes~\cite{10.1145/3313831.3376293} that can be used to determine whether a news source is reliable or not.

Our findings are consistent with Jahanbakhsh et al., who showed that people believe in a news article if a claim is from a source they trust~\cite{farnaz2021lightweight}. \added{If the provider of the source label is perceived as reliable and trustworthy, the provider of the label could act as a source that a user can trust}. This aligns with early work by Pornpitakpan, who showed that a message from a highly credible source is more likely to be believed than the same message published by a source with low credibility~\cite{pornpitakpan2004persuasiveness}. The results are also consistent with recent findings by Arnold et al., \final{who found that source labels can reduce belief in disinformation claims and users' sharing intentions of disinformation}~\cite{arnold2021source}.

Interventions like the Assess the Source component could support users in deciding whether something is unreliable or not. The advantage of this approach is that source labels are easier to implement than tools that provide evidence for or against a claim~\cite{pennycook2018prior} or that try to change how users perceive author names, domain names, topics, and logos~\cite{morris2012tweeting,wineburg2017lateral}. Our findings corroborate Kirchner and Reuter, who showed that warning labels can effectively reduce the perceived accuracy of false headlines, at least in Germany~\cite{10.1145/3415211}. In our investigation, participants had access to the entire website, i.e., they could engage deeply with the content. We, therefore, extend on prior work focused on headlines and content labels with a focus on full articles and source labels. 
Our investigation also indicates that there are important differences between the countries and cultures. \added{O}ur results pose the question of how well Kirchner and Reuter's results in Germany can be generalized to other countries like the United States. We find that there are cultural differences that need to be taken into account. This is especially important because social media platforms like Facebook, YouTube, Telegram, and TikTok have users from all over the world. We encourage further research on why source labels did not lead to better article ratings in the U.S. even though they were perceived as helpful.

While our results demonstrate the potential of source labels, it is important to consider their risks. History proved that even reliable sources are involved in controversies. For instance, Judith Miller at the New York Times falsely reported that Saddam Hussein had or was acquiring weapons of mass destruction, which influenced the decision to invade Iraq~\cite{enwiki:1032880882}. More recently, German ``reporter'' Claas Relotius was found guilty of fabricating at least 14 stories~\cite{enwiki:1020503053}. This means that reliable sources are not infallible. Considering how rare such instances are and how much public outrage they caused, we still believe that source labels are a quick and powerful approximation that can help users in the vast majority of cases. Another key benefit of source labels is that---unlike content labels---they can account for possible but unlikely events that need to be reported and which a content-based \added{machine learning} approach might flag as false. For example, the secret global surveillance programs by the National Security Agency (NSA) were considered to be conspiracy theories until Edward Snowden provided evidence. We, therefore, conclude that source labels are a very useful direction in the fight against misinformation. \final{Considering the possible influence of order effects, we invite further research that examines the impact of source labels in more detail.}

\subsection{Information Landscapes and Their Possible Impact}

\final{In the context of the ordered checklist provided by the WHO, we found} that source labels may be preferable to interventions that require more user engagement, e.g., doing research on the content or author of an article, even when the interface is designed to lower the cost of doing that research. One caveat is that future novel interface designs may reverse this effect. \added{Prior research showed that distinguishing reliable from unreliable news is a task that most people are capable of doing~\cite{pennycook2019crowdsourcing,pennycook2019lazy,10.1145/3240167.3240172}. In our investigation, we were, therefore,} surprised about the low number of correct article reliability ratings in both countries and the strong differences between Germany and the United States. 
For example, in the Interactive Checklist condition, the source label always correctly reflected the reliability of an article. \added{If a participant would have followed the recommendation by the 1.~Assess the Source component of the Interactive Checklist, he or she would have gotten all three article ratings right. Yet, only 26\% of Germans and 19\% of Americans got all three ratings right.} %

\added{Our results show that} the article ratings in Germany did improve with the introduction of both the Written and Interactive Checklists, but we did not see an improvement in the United States with either intervention (Figure~\ref{fig:correct_article_ratings}).
The results in the United States are consistent with recent findings that suggest that people may stick to their initial decisions, regardless of the reasons that are presented to them~\cite{stanley2020resistance,kahne2017educating}. This prior-belief bias could be the explanation for why US participants did not change their opinions but the bias does not explain why German participants benefited from the interventions. This \added{difference} may be a consequence of the different information landscapes in Germany and the United States. According to the Reuters Digital News Report 2021, half of the respondents (53\%) in Germany state that they have an overall trust in news~\cite{newman2021reuters}. In the United States, only 29\% state that they have an overall trust in news---the worst of all 46 countries covered by the report. This is especially noteworthy because trust in social media for news (DE: 14\%, US: 13\%) and trust in search engines for news (DE: 25\%, US: 22\%) \final{is} similar. The differences in participants' underlying trust in media, and by extension, trust in our interventions about that media, could have influenced the results. \added{Our findings are consistent with how citizens from the different countries self-report their concern about what is real and what is fake on the Internet. In Germany, only 37\% of respondents in the Reuters Digital News Report 2020 state that they are concerned about what is real and what is fake~\cite{newman_overview_2020}. In the United States, two out of three (67\%) of the respondents are concerned. Our results corroborate these concerns. Participants from Germany, where people are less concerned, classified more articles correctly than participants from the United States, where the concern is higher.}

The degree of polarization may compound the differences in trust in media seen between these two countries. In Germany, the top news services are public service broadcasters like ARD (54\%) and ZDF (45\%)~\cite{newman2021reuters}. German public broadcasters are independent of the government and required by law to be ``independent and objective''~\cite{interstate_broadcasting_agreement}. German law demands that ``news must be verified regarding their truthfulness and origin in accordance with the attention to accuracy and source required by the circumstances''. To achieve this, public broadcasters are supervised by so-called ``broadcasting councils'', which were implemented after World War II to limit state influence on public broadcasting in Germany~\cite{fr_ww2_2009}. In these councils, representatives of societally relevant groups ensure the independence and objectivity of the broadcasters~\cite{kleinsteuber2011control}. A 2018 poll by Pew Research Center showed that the German public broadcaster ARD is by far the most trusted news source by both people from the left and the right~\cite{pew_germany_media_trust_2018}. \added{Trust in these public broadcasters is higher than in any other German outlet.} In the United States, the top brands are less popular and more biased. Besides local TV news (26\%), Fox News (25\%) and CNN (24\%) are the most \final{trusted} news sources. \added{The meta-ranking by Gruppi et al.~labels both as biased~\cite{gruppi2020nelagt2019}. Fox News is labeled as biased towards the right. CNN is labeled as biased towards the left.} 
This comparison could imply that people in the United States are more used to encountering biased and potentially unreliable reporting, which could have influenced their ratings of news articles. One consequence of \added{a biased information landscape} could be that users become better at distinguishing unreliable from reliable news articles. However, for the limited sample of news articles that we investigated, this was not the case. Another possible interpretation, especially considering the reliance on biased sources in the United States, could be an increasing commodification of truth. People may not seek an intersubjective agreement~\cite{10.5840/monist19735722} on facts, but rather see information like just another market in which they can pick their preference. This, of course, is problematic, because people might end up in a situation where they turn to biased or unreliable sources to ``fact-check'' information. This could explain why the kind of fact-checking promoted by the checklist by the WHO may not be effective in the U.S. This explanation connects to boyd, who warned that due to polarization, distrust, and self-segregation, the United States is ``moving towards tribalism''~\cite{boyd_2018_b}. This could likewise explain why the interventions were more effective in Germany, where polarization and self-segregation are comparatively low and where trust is high, especially in the public broadcasting services~\cite{pew_germany_media_trust_2018,newman2021reuters}. \final{In} Germany, the checklists led to better performance at the article rating tasks, although this performance is still far from perfect. 

Overall, our investigation showed that both the Written and Interactive Checklists may be most effective in information landscapes with high media trust because they leverage trust in sources to explain whether something is reliable or not. In an information environment with low media trust, this approach may be less helpful because users may rely on other means to assess the truth value of information. This relates to boyd's experience with teens who---following the recommendation that Wikipedia is untrustworthy and that they had to do their own ``research''---identified websites online that ``proved'' their beliefs~\cite{boyd_2018_b}. Boyd believes that ``addressing so-called fake news is going to require a lot more than labeling''~\cite{boyd_2018_b}. She argues that it is ``going to require a cultural change about how we make sense of information, whom we trust, and how we understand our own role in grappling with information''. She also warns that quick and easy solutions will not address the underlying problems. We contribute towards this endeavor by empirically validating interventions proposed by the WHO for the United States and Germany. While we acknowledge boyd's criticism, we find that \final{the ordered checklist is effective, especially in Germany, and that} the source labels in our paper outperformed the other approaches \final{in this setting}.

\subsection{Design Recommendations}

Our investigation showed that a large subset of users appreciates help with distinguishing reliable from unreliable information. \added{Our findings also show that technical tools can help participants distinguish between reliable and unreliable information.} We also found that not every group of users benefits equally. %
This, of course, poses the question of how the intervention can be improved to support \final{everybody}. \added{The linear mixed model analysis indicated that age has a small, but significant influence on participants' ratings. Our analysis showed that especially} young adults in the United States are a subgroup that requires special attention. As our investigation showed, the task performance at rating news articles by young adults in the U.S.~is significantly worse than the performance of middle-aged people, who in turn perform significantly worse than elderly people. We invite further research to examine why the ratings of young adults deviated from the other groups so strongly in our investigation in the United States, especially since we did not observe a similar trend in Germany.

\added{Our correlation analysis found that acting on a component influences the perceived helpfulness of the component. We find that the more people act on a certain component, the more helpful they rate this component. Those components that are perceived as helpfulness and acted on the most frequently also have the strongest effect on task performance.}

Considering the benefits of source labels that we discussed in the previous sections, we encourage providers of information systems to add sources labels that mark \textit{all} news sources as reliable or unreliable to mitigate the possible impact of the ``Implied Truth Effect''~\cite{pennycook2020implied}. Marking all news sources prevents that users assume that sources without a label are validated and accurate. The reliability of a source can also be explained without the user directly engaging with the misinformation, which limits the risk of backfire effects~\cite{wood2019elusive}. Explanations for the unreliability of a source could just point out the number of misinformation stories published by the source~\cite{farnaz2021lightweight}. Therefore, our recommendation is to add democratically and transparently determined source labels to online platforms, e.g., based on the digital juries approach~\cite{10.1145/3313831.3376293}. Such labels could be easily integrated into the user interface of social media platforms like Facebook, YouTube, TikTok, and Twitter as well as messenger services like WhatsApp and Telegram. As detailed, this could help users, especially in Germany, quickly assess whether a news story is reliable or not. 

\subsection{Limitations \& Future Work}

It is important to note that our recommendations are influenced by our positionality as academics born and raised in Germany and the United States, respectively. Our internal discussion showed that even the German and the U.S.~perspectives, e.g., in relation to the utility of source labels, differ. Our findings are limited to the field sites that we selected. We focused on the Western, democratic countries in which we grew up and currently live. While the two countries that we compared are very different in regards to their trust in media, it would also be interesting to investigate the perception of misinformation in other countries, e.g., from Asia, Africa, or South America. Further research is necessary to show how applicable our findings are in other cultural contexts. 

Studying which interventions are actually helpful in the fight against misinformation is challenging. \added{In our study, participants were in a familiar environment and used their own computer, but the articles they viewed were not selected by themselves or shared by their social network}. And while they may not have had to explicitly rate the reliability of news articles before, they have to implicitly do this many times every day. 

We relied on Lucid, a professional market research company, to recruit a gender-balanced sample that is diverse in regards to the political opinions, education, and age of users. Prior research showed that subjects recruited from the Lucid platform constitute a sample that is suitable for evaluating many social scientific theories~\cite{doi:10.1177/2053168018822174}. Nevertheless, it is possible that we were not able to reach those who are most disenfranchised from contemporary information landscapes. We invite other researchers to publish more research on who would benefit from interventions against misinformation the most and how to reach them.

\added{For an investigation like this, it is important to take the possible influence of order effects into account because the order of the \final{checklist} components could have influenced how they were perceived. We decided against modifying the order of the components because we wanted to investigate the checklist of the World Health Organization in a realistic experiment using real news \final{articles}. Our findings are representative of the checklist provided by the WHO that was used during the COVID-19 pandemic.} \final{Since we did not randomize the order of the different checklist components, we cannot rule out that the ranking of the components by a global authority like the WHO influenced the perception of the components. We, therefore, encourage further research to study the order effects of checklists against misinformation. Our results indicate that the WHO successfully prioritized the different components in the checklist, e.g., to reduce the potential negative effect of banner blindness or user fatigue~\cite{hervet2011banner,doi:10.1177/154193129804200504}.}

\added{Another notable finding of our investigation is that young adults in the U.S. and in Germany are worse at identifying reliable news than other age groups. The empirical data that we collected did not allow us to explain why young adults' task performance is so limited. Considering the possible implications of this finding, we, therefore, invite other researchers to examine these differences in future work.}

\section{Conclusion}

\added{This paper empirically shows that users can be supported in the fight against misinformation, especially in Germany. 
With this paper, we provide the first empirical investigation of the written checklist provided by the WHO. We also show how interactive elements can increase the effectiveness of the checklist. For this}, we augmented and evaluated checklists based on recommendations by the WHO. 
We find that both the Written Checklist and the Interactive Checklist are perceived as helpful by users, especially the source labels. 
In Germany, we also find that the Interactive Checklist significantly improves users' performance at the article rating task. \added{Acting on the recommendation to assess the source is correlated with better task performance.} While participants in the U.S. perceive the checklist as helpful, we do not find a measurable difference in their article ratings. We relate these differences between Germany and the U.S. to the different information landscapes and the differences in trust in media. Based on our insights, we make concrete design recommendations. We believe that since source labels are perceived favorably by users, they are \final{a} promising direction in the fight against misinformation, especially considering the limitations that we identified for fact-checks\final{, even though further research is needed on the effect of the order in which components are presented in checklists against misinformation}. We hope that the insights presented in this paper motivate social media providers as well as civic hackers to develop tools that support users in distinguishing unreliable from reliable information.

\begin{acks}
This research has been supported by the German Research Foundation, Grant No.~374666841 (SFB 1342), and the NSF under Grant No.~2107391. We thank all participants for their time and their insights. We also thank the reviewers for their exceptionally helpful and constructive feedback. Special thanks go to Andreas Breiter, Nils Düpont, Martin Scharpenberg, Gabriela Molina León, and Michael Windzio for their help and advice. Statistical support was provided by data science specialist Steven Worthington, at the Institute for Quantitative Social Science, Harvard University.
\end{acks}

\bibliographystyle{ACM-Reference-Format}
\bibliography{references-experiment}

\appendix

\section{Further Correlation Analysis}

\final{We investigated the correlations between acting on different components, between perceiving different components as helpful, and between acting on a component and perceived helpfulness using Spearman's $\rho$. We report these correlations for the German study in the top left of Table~\ref{tab:correlations_germany} and the results for the U.S. study in the top left of Table~\ref{tab:correlations_usa} that you find on the next page.}

\begin{table*}[t]
\small
\centering
\caption{\added{In the German study, Spearman's $\rho$ identifies correlations between acting on different components (Act, Figure~\ref{fig:recommendations_acted}) and the perceived helpfulness of the components (Help, Figure~\ref{fig:recommendations_usefulness}). Whether somebody acts on a component correlates with whether he or she perceives a component as helpful. Acting on any of the components also correlates with an increase in the perceived helpfulness of the 1.~Assess the Source component. *~$\leq$~.05, **~$\leq$~.01, ***~$\leq$~.001.}}
\label{tab:correlations_germany}
\Description{This table identifies correlations between acting on different components and the perceived helpfulness of the components for the German study based on Spearman's Rho. We find that whether somebody acts on a component correlates with whether he or she perceives a component as helpful. Acting on any of the components also correlates with an increase in the perceived helpfulness of the 1. Assess the Source component.}
\begin{tabular}{p{3.2cm}|p{0.7cm}p{0.8cm}p{0.7cm}p{0.7cm}p{0.8cm}p{0.8cm}p{0.9cm}|p{0.8cm}p{0.8cm}p{0.7cm}p{0.7cm}}
\toprule
& \multicolumn{7}{c|}{Acted on Component (Act)} & \multicolumn{4}{c}{Helpfulness of Components (Help)} \\
$corr(x,y)$ & Assess Source & Headline & Author & Date & Evidence & Fact-Checking & Fact-Checkers & Assess Source & Headline & Author & Date \\
\midrule
Act: Beyond Headline & .43 *** &  &  &  &  &  &  &  &  &  &  \\
Act: Identify Author & .40 *** & .44 *** &  &  &  &  &  &  &  &  &  \\
Act: Check Date & .36 *** & .32 *** & .26 *** &  &  &  &  &  &  &  &  \\
Act: Evidence & .46 *** & .40 *** & .48 *** & .33 *** &  &  &  &  &  &  &  \\
Act: Fact-Checking & .37 *** & .42 *** & .33 *** & .29 *** & .57 *** &  &  &  &  &  &  \\
Act: Turn to Fact-Checkers & .19 * & .29 *** & .36 *** & .35 *** & .47 *** & .44 *** &  &  &  &  &  \\
\midrule
Help: Assess Source & .34 *** & .16 * & .21 ** & .17 * & .20 ** & .22 ** & .15 * &  &  &  &  \\
Help: Beyond Headline & .17 * & .21 ** & .09 & .13 & .08 & .02 & .11 & .19 ** &  &  &  \\
Help: Identify Author & .23 ** & .23 ** & .33 *** & .19 ** & .15 * & .24 *** & .20 ** & .37 *** & .36 *** &  &  \\
Help: Check Date & .13 & .13 & .14 & .33 *** & .13 & .16 * & .12 & .31 *** & .32 *** & .37 *** &  \\
Help: Turn to Fact-Checkers & .19 * & .05 & .13 & .19 * & .19 ** & .18 * & .18 * & .22 ** & .41 *** & .30 *** & .37 *** \\
\bottomrule
\end{tabular}
\end{table*}

\begin{table*}[t]
\small
\centering
\caption{\added{Regarding correlations between acting on components (Act) and perceiving components as helpful (Help), the same trends can be observed in the U.S. study. While Spearman's $\rho$ identifies fewer and weaker correlations between acting on different components (Act, Figure~\ref{fig:recommendations_acted}) and the perceived helpfulness of the components (Help, Figure~\ref{fig:recommendations_usefulness}) than the study in Germany (Table~\ref{tab:correlations_germany}), the same general tendencies can be observed, including the observation that acting on any component increases the perceived helpfulness of the 1. Assess the Source component. *~$\leq$~.05, **~$\leq$~.01, ***~$\leq$~.001.}}
\label{tab:correlations_usa}
\Description{This table identifies correlations between acting on different components and the perceived helpfulness of the components for the U.S. study based on Spearman's Rho. While we identify fewer and weaker correlations between acting on different components and the perceived helpfulness of the components than the study in Germany, the same general tendencies can be observed, including the observation that acting on any component increases the perceived helpfulness of the 1. Assess the Source component.}
\begin{tabular}{p{3.2cm}|p{0.7cm}p{0.8cm}p{0.7cm}p{0.7cm}p{0.8cm}p{0.8cm}p{0.9cm}|p{0.8cm}p{0.8cm}p{0.7cm}p{0.7cm}}
\toprule
& \multicolumn{7}{c|}{Acted on Component (Act)} & \multicolumn{4}{c}{Helpfulness of Components (Help)} \\
$corr(x,y)$ & Assess Source & Headline & Author & Date & Evidence & Fact-Checking & Fact-Checkers & Assess Source & Headline & Author & Date \\
\midrule
Act: Beyond Headline & .45 *** &  &  &  &  &  &  &  &  &  &  \\
Act: Identify Author & .34 *** & .35 *** &  &  &  &  &  &  &  &  &  \\
Act: Check Date & .27 *** & .38 *** & .46 *** &  &  &  &  &  &  &  &  \\
Act: Evidence & .42 *** & .52 *** & .36 *** & .47 *** &  &  &  &  &  &  &  \\
Act: Fact-Checking & .29 *** & .52 *** & .37 *** & .43 *** & .55 *** &  &  &  &  &  &  \\
Act: Turn to Fact-Checkers & .31 *** & .42 *** & .34 *** & .48 *** & .50 *** & .52 *** &  &  &  &  &  \\
\midrule
Help: Assess Source & .23 *** & .24 *** & .22 ** & .24 *** & .17 * & .20 ** & .23 ** &  &  &  &  \\
Help: Beyond Headline & .13 & .17 * & .23 ** & .12 & .11 & .13 & .06 & .10 &  &  &  \\
Help: Identify Author & .11 & .12 & .37 *** & .2 ** & .07 & .13 & .19 ** & .29 *** & .28 *** &  &  \\
Help: Check Date & .01 & .03 & .08 & .17 * & .05 & .07 & .13 & .11 & .17 * & .32 *** &  \\
Help: Turn to Fact-Checkers & .07 & .12 & .15 * & .03 & .16 * & .13 & .22 ** & .08 & .19 ** & .22 ** & .33 ***\\
\bottomrule
\end{tabular}
\end{table*}

\end{document}